# Experimental demonstration of electric power generation from Earth's rotation through its own magnetic field

Christopher F. Chyba [*]
*Department of Astrophysical Sciences and School of Public and International Affairs, Princeton University, Princeton, New Jersey 08544, USA*

Kevin P. Hand
*Jet Propulsion Laboratory, California Institute of Technology, Pasadena, California 91109, USA*

Thomas H. Chyba
*Spectral Sensor Solutions, LLC, Albuquerque, New Mexico 87111, USA*

Earth rotates through the axisymmetric part of its own magnetic field, but a simple proof shows that it is impossible to use this to generate electricity in a conductor rotating with Earth. However, we previously identified implicit assumptions underlying this proof and showed theoretically that these could be violated and the proof circumvented. This requires using a soft magnetic material with a topology satisfying a particular mathematical condition and a composition and scale favoring magnetic diffusion, i.e., having a low magnetic Reynolds number $R_m$ [Chyba and Hand, Phys. Rev. Appl. **6**, 014017 (2016)]. Here we realize these requirements with a cylindrical shell of manganese-zinc ferrite. Controlling for thermoelectric and other potentially confounding effects (including 60 Hz and RF background), we show that this small demonstration system generates a continuous DC voltage and current of the (low) predicted magnitude. We test and verify other predictions of the theory: voltage and current peak when the cylindrical shell's long axis is orthogonal to both Earth's rotational velocity **v** and magnetic field; voltage and current go to zero when the entire apparatus (cylindrical shell together with current leads and multimeters) is rotated 90° to orient the shell parallel to **v**; voltage and current again reach a maximum but of opposite sign when the apparatus is rotated a further 90°; an otherwise-identical *solid* MnZn ferrite cylinder generates zero voltage at all orientations; and a high-$R_m$ cylindrical shell produces zero voltage. We also reproduce the effect at a second experimental location. The purpose of these experiments was to test the existence of the predicted effect. Ways in which this effect might be scaled to generate higher voltage and current may now be investigated.

## I. INTRODUCTION

Could electricity be generated from Earth's rotation through its own magnetic field? This question has been asked at least since Faraday's first experiments testing the idea in January 1832 gave a negative result [1], and the answer, for reasons reviewed below, has since seemed to remain obviously no [2].

But in 2016 we showed theoretically that for a system satisfying specific topological and material conditions the answer could be yes [2]. Here we present experimental results for a small demonstration laboratory system that generates a low continuous DC voltage and current that behave according to that prediction. The intention of these experiments was to test the existence of the predicted effect, and the results and multiple controls we report here appear to demonstrate its reality. Ways in which this effect might be scaled to generate higher voltage and current are proposed but will be the subject of subsequent investigations.

We first briefly discuss historical experiments relevant to this question and cast these results in a modern understanding of electromagnetism while providing a few necessary definitions. We then review the physics that, surprisingly, makes the generation of electricity from Earth's rotation through its own field possible. In the subsequent section we determine the conditions required for a system to generate a continuous DC voltage as Earth carries that system through its magnetic field [2]. We derive specific predictions for this system (both in voltage and current magnitude and behavior under rotation), describe our experimental materials and methods, and compare our experimental results with the predictions. We include a discussion of controls and how we excluded possible confounding effects. Finally, we review previous experiments and objections, and conclude with a survey of ways this effect might be scaled to generate higher voltage and current.

[*]Contact author: cchyba@princeton.edu

### A. Earth's geomagnetic field

Throughout this paper, we will be concerned with conductors rotating with Earth's surface either through or along with components of Earth's magnetic field. To establish notation for this discussion, consider two reference frames, $K$ and $K'$. $K$ is an inertial frame with origin at Earth's center with the usual spherical coordinates $(r, \theta, \varphi)$; $K$ moves with Earth in its orbit but does not rotate with the planet's polar ($z$-axis) rotation. The origins of $K'$ and $K$ coincide, but $K'$ corotates with Earth at angular frequency $\boldsymbol{\omega} = \omega\hat{z}$, so that a point fixed on Earth's surface does not change its coordinates in $K'$ over time. $K'$ is therefore the laboratory frame, and has coordinates $(r, \theta, \varphi')$ where $\varphi' = \varphi - \omega t$ for time $t$. Frames $K$ and $K'$ are not exactly related by a Lorentz transformation because of Earth's rotation. In $K'$, Maxwell's equations incorporate rotation via the metric tensor $g_{\mu\nu}$, introducing factors $\sqrt{g_{00}} \approx 1 - \frac{1}{2}(v/c)^2$ when $(v/c) \ll 1$, where $c$ is the speed of light [3]. At velocities relevant to Earth's rotation ($v = 354\,\mathrm{m\,s^{-1}}$ at Princeton's latitude, where our experiments were performed), electromagnetism in $K'$ behaves like that in an inertial frame to $\mathcal{O}(v/c)^2 \sim 10^{-12}$. These corrections are negligible compared to the effects of interest here [2]. We may therefore approximate $K$ and $K'$ as two inertial frames in relative linear motion. We henceforth assume $(v/c)^2 \ll 1$ throughout this paper.

Detailed models of Earth's field derive it from a magnetic potential written in terms of surface harmonics and Schmidt-normalized associated Legendre polynomials with coefficients $g_l^m$ and $h_l^m$. A contemporary model carries terms up through degree $l = 13$ and order $\mathrm{m} = 12$ [4]. Earth's tilted magnetic dipole can be resolved into components symmetric about Earth's rotation axis, and nonaxisymmetric components that depend on $\cos \mathrm{m}\varphi'$ and $\sin \mathrm{m}\varphi'$, where $\varphi'$ is the azimuthal angle (longitude) in $K'$ and m a whole number [4–7]. The nonaxisymmetric components have order $\mathrm{m} \geqslant 1$, whereas the symmetric components are of order $\mathrm{m} = 0$. In particular, the $g_1^0$ term corresponds to Earth's primary dipole axisymmetric about (and antiparallel to) Earth's rotation axis, and the leading off-axis terms $g_1^1$ and $h_1^1$ correspond to weaker dipoles lying 90° apart in Earth's equatorial plane and rotating with the planet [6].

Earth's axially symmetric dipole is given by

$$\begin{aligned} B_r^{\mathrm{m}=0} &= 2g_1^0(a/r)^3\cos\theta, \\ B_\theta^{\mathrm{m}=0} &= g_1^0(a/r)^3\sin\theta, \\ B_\varphi^{\mathrm{m}=0} &= 0, \end{aligned} \tag{1}$$

where $g_1^0 = -24496.5\,\mathrm{nT}$ [4], and the superscript "$\mathrm{m} = 0$" labels these as components of the axisymmetric field. Quadrupole, octupole, and higher degree axisymmetric terms enter with coefficients $g_2^0 = -2396.6\,\mathrm{nT}$, $g_3^0 = 1339.7\,\mathrm{nT}$, and so on. Obviously, $\mathbf{B}^{\mathrm{m}=0}$ has no $\varphi$ dependence for any degree $l$.

The lowest-degree nonaxisymmetric field is given by

$$\begin{aligned} B_r^{\mathrm{m}\neq0} &= 2(a/r)^3\left(g_1^1\cos\varphi' + h_1^1\sin\varphi'\right)\sin\theta, \\ B_\theta^{\mathrm{m}\neq0} &= -(a/r)^3\left(g_1^1\cos\varphi' + h_1^1\sin\varphi'\right)\cos\theta, \\ B_\varphi^{\mathrm{m}\neq0} &= (a/r)^3\left(g_1^1\sin\varphi' - h_1^1\cos\varphi'\right), \end{aligned} \tag{2}$$

where $g_1^1 = -1585.9\,\mathrm{nT}$, $h_1^1 = 4945.1\,\mathrm{nT}$ [4], and the superscript "$\mathrm{m} \neq 0$" labels these as components of the nonaxisymmetric field. Considering Eq. (2) in the $\theta = 90°$ (equatorial) plane makes it evident that the $g_1^1$ and $h_1^1$ terms have the form of dipoles oriented along the $\hat{x}'$ and $\hat{y}'$ axes. Because $\varphi'$ denotes a longitude in $K'$, $\mathbf{B}^{\mathrm{m}\neq0}$ rotates in $K$ with Earth at angular speed $\omega$.

### B. Historical background and definitions

The behavior of conductors rotating through magnetic fields was the subject of considerable investigation during the nineteenth and early twentieth centuries [8,9]. In 1852, Faraday experimented with a rotating conducting magnet connected to a galvanometer via contacts on the magnet's axle and rim [10]. Current flowed when the magnet rotated around its north-south axis and the galvanometer part of the circuit $C$ remained stationary, or when the magnet was stationary but the circuit rotated.

In modern terms, the conducting magnet rotates at velocity $\mathbf{v}$ through its own magnetic field $\mathbf{H}$ (where magnetic flux density $\mathbf{B} = \mu\mathbf{H}$ and $\mu$ is magnetic permeability), producing a $\mathbf{v} \times \mathbf{B}$ field that generates an electromotive force (emf), driving a current. The emf $\varepsilon$ around a path $C$ with line element $\mathbf{dl}$ moving with velocity $\mathbf{v}$ is [11]:

$$\begin{aligned} \varepsilon &= \oint_C (\mathbf{E} + \mathbf{v} \times \mathbf{B}) \cdot \mathbf{dl} \\ &= \int_S [-\partial\mathbf{B}/\partial t + \nabla \times (\mathbf{v} \times \mathbf{B})] \cdot \mathbf{da}, \end{aligned} \tag{3}$$

where $\mathbf{E}$ is the electric field, $\mathbf{B}$ the magnetic flux density, and the area element $\mathbf{da}$ is right-hand normal to the surface $S$ bounded by $C$. The second equality in Eq. (3) holds via Stokes' theorem provided there is no jump discontinuity on $S$ [12]. This will not be at issue for our experiments below. The emf in $K'$ is the same as that in $K$ provided that $(v/c)^2 \ll 1$ [13].

Faraday interpreted his results to mean that magnetic fields do not rotate with a magnet when the magnet rotates around its symmetry axis [10]. But Preston [14] showed that Faraday's results were equally well explained in a picture where the magnetic field *does* rotate with the magnet, and thereby produces a $q\mathbf{v} \times \mathbf{B}$ force on charges $q$ in the stationary part of $C$, were $\mathbf{v}$ understood to mean the velocity of the rotating field. At the turn of the century, Poincaré concluded that since both the rotating and nonrotating field pictures appeared to give identical results, the distinction between them was "meaningless" [15].

Nonetheless, in 1912 Barnett [16] reported open-circuit experiments that settled the question. Barnett placed a cylindrical capacitor axially in the field of a solenoid (or, in another version of the experiment, in the field between two electromagnets capped with flat pole pieces), with a wire connecting the concentric cylinders of the capacitor. Corotation of the nested cylinders and their connecting wire charges the cylinders due to the $q\mathbf{v}\times\mathbf{B}$ force on the wire; the wire may then be disconnected, the system despun, and an opposite charge on the cylinders measured. But Barnett showed that rotating the solenoid (or electromagnets) while holding the capacitor and connecting wire stationary did not charge the cylinders. These results were reproduced by Kennard [17], and Pegram showed that corotation of the capacitor and connecting wire together with the solenoid also charged the capacitor [18]. These results proved that the field of a rotating axially symmetric electromagnet does not "rotate with the magnet," a conclusion that is now a standard part of the electromagnetics literature [3,19–21]. A subsequent attempted theoretical refutation of these results was shown to be incorrect due to calculational error [22–24].

Some authors have nonetheless argued that an axisymmetric field rotates with a *permanent* magnet [25]. But even if this conclusion were correct, a permanent magnet cannot provide a model for Earth's deeply originating magnetic field, which is the field that concerns us here. Earth's field derives from an electromagnetic dynamo in Earth's liquid-iron outer core, which is at a depth of more than 2800 km [5,6]. Iron's Curie temperature is reached at only ~30 km depth [5], and Earth's magnetic field reversals are also inconsistent with a permanent magnet [6].

In the usual modern understanding of the Lorentz force $\mathbf{F}=q(\mathbf{E}+\mathbf{v}\times\mathbf{B})$, the only relevant velocity $\mathbf{v}$ is the velocity of a charge $q$ (or of a conductor, which provides a collection of charges) relative to a frame in which the magnetic flux density is $\mathbf{B}$ [26]. The notion of a uniform field "moving" or "rotating" is regarded as incoherent [27]. Nonuniform fields can of course translate (for example, a field displacing along with the moving magnet that generates it) or rotate, as with the m $\neq 0$ components of Earth's geomagnetic field, Eq. (2). This may introduce a time-varying $\mathbf{B}$ field, $\partial\mathbf{B}/\partial t$, in a particular frame, or correspondingly a term $\mathbf{E}=-\partial\mathbf{A}/\partial t$ in the Lorentz force law, but the velocity $\mathbf{v}$ in the $\mathbf{v}\times\mathbf{B}$ part of that law refers only to the velocity of the charge $q$ in the frame under consideration, not a velocity relative to the translating or rotating magnetic field.

### C. Lorentz force for m = 0 and m ≠ 0 components of Earth's field

Consider a test charge $q$ corotating with Earth at a constant velocity $\mathbf{v}=r\omega\sin\theta\hat{\varphi}$ as viewed in reference frame $K$. We first consider how this test particle interacts with the axisymmetric part $\mathbf{B}^{\mathrm{m}=0}$ of Earth's field, the leading term of which is the dipole field explicitly displayed in Eq. (1). In $K$, $-\partial\mathbf{B}^{\mathrm{m}=0}/\partial t=\mathbf{0}=\nabla\times\mathbf{E}$. The condition $\nabla\times\mathbf{E}=\mathbf{0}$ allows $\mathbf{E}=-\nabla V$, with $V$ an electric potential, but absent any electrostatic charge buildup (we will return to this below), $\nabla V=\mathbf{0}$ and we put $\mathbf{E}=\mathbf{0}$ in the Lorentz force law so that $q$ experiences a force

$$\mathbf{F}^{\mathrm{m}=0}=q\mathbf{v}\times\mathbf{B}^{\mathrm{m}=0}. \tag{4}$$

This is the Earth system's manifestation of the Lorentz force observed in the Barnett experiments. The force $\mathbf{F}^{\mathrm{m}=0}$ must be the same force to $\mathcal{O}(v/c)^2$ in $K'$ [3]. In frame $K'$ (our laboratory frame), $q$ has $\mathbf{v}=\mathbf{0}$ relative to the frame, so there is no $\mathbf{v}\times\mathbf{B}$ force. However, there is an electric field in $K'$ due to the relativistic field transformation to $O(v/c)^2$:

$$\mathbf{E}'=\mathbf{E}+\mathbf{v}\times\mathbf{B}, \tag{5}$$

and the Lorentz force law in $K'$ becomes, since $\mathbf{E}=\mathbf{0}$,

$$\mathbf{F}'^{\mathrm{m}=0}=q\mathbf{E}'=q(\mathbf{E}+\mathbf{v}\times\mathbf{B}^{\mathrm{m}=0})=q\mathbf{v}\times\mathbf{B}^{\mathrm{m}=0}. \tag{6}$$

That is, in $K'$ there is no $\mathbf{v}\times\mathbf{B}$ force since $\mathbf{v}=\mathbf{0}$ but there is now an electric field $\mathbf{E}'$ that exerts a force on $q$ identical to that due to $\mathbf{v}\times\mathbf{B}$ in $K$. We note that the field transformation to $O(v/c)^2$ for the magnetic flux density is just $\mathbf{B}'=\mathbf{B}$.

Does the charge $q$ also experience a net force from its motion in frame $K$ in the presence of the $\mathbf{B}^{\mathrm{m}\neq 0}$ field? The leading components of the $m\neq 0$ field are given by Eq. (2). In a modern understanding of $\mathbf{v}$ in the Lorentz force law, the fact that these components are also, like $q$, rotating at frequency $\omega$ in $K$ does not change the existence of a $q\mathbf{v}\times\mathbf{B}^{\mathrm{m}\neq 0}$ force on $q$. But for the m $\neq 0$ components there is an additional force present that is absent for the m $=0$ components, because $\partial\mathbf{B}^{\mathrm{m}\neq 0}/\partial t\neq\mathbf{0}$ in $K$. In frame $K$ Maxwell's Faraday equation requires

$$\partial\mathbf{B}^{\mathrm{m}\neq 0}/\partial t=-\nabla\times\mathbf{E}. \tag{7}$$

By direct calculation from Eq. (2) using $\varphi'=\varphi-\omega t$, it is easy to verify

$$\partial\mathbf{B}^{\mathrm{m}\neq 0}/\partial t=\nabla\times(\mathbf{v}\times\mathbf{B}^{\mathrm{m}\neq 0})=-\omega\partial\mathbf{B}^{m\neq 0}/\partial\varphi'\neq\mathbf{0}, \tag{8}$$

as expected for a translating system with no diffusion. Equating Eqs. (7) and (8), the simplest choice for $\mathbf{E}$ is

$$\mathbf{E}=-\mathbf{v}\times\mathbf{B}^{\mathrm{m}\neq 0}. \tag{9}$$

So in frame $K$ for the m $\neq 0$ components there is a $\mathbf{v}\times\mathbf{B}$ field but also an $\mathbf{E}$ field of equal and opposite sign. The Lorentz force in $K$ due to the m $\neq 0$ components is then

$$\mathbf{F}^{m\neq 0}=q(\mathbf{E}+\mathbf{v}\times\mathbf{B}^{\mathrm{m}\neq 0})=\mathbf{0}, \tag{10}$$

which means $\mathbf{F}^{\mathrm{m}\neq 0}=\mathbf{0}$ in $K'$ as well.

Why would a charge $q$ feel no force from the m $\neq 0$ components in the laboratory frame? As with the m $=0$ components, $q$ has $\mathbf{v}=\mathbf{0}$ in the laboratory, so there is no $\mathbf{v}\times\mathbf{B}$ force. But by Eqs. (5) and (9), the electric field $\mathbf{E}'=\mathbf{0}$ in $K'$, unlike the case for the m $=0$ components. Then Ohm's law in $K'$ gives

$$\mathbf{F}'^{\mathrm{m}\neq 0}=q\mathbf{E}'=\mathbf{0}. \tag{11}$$

Formally Eqs. (7) and (8) admit a solution $\mathbf{E}=-\mathbf{v}\times\mathbf{B}^{\mathrm{m}\neq 0}-\nabla V$, which would give $\mathbf{E}'=-\nabla V$. (We note $\nabla=\nabla'$.) But this $V$ cannot establish a Lorentz force for the m $\neq 0$ components analogous to Eq. (4). By Eq. (11), if we hoped to achieve $\mathbf{F}'=q\mathbf{v}\times\mathbf{B}^{\mathrm{m}\neq 0}$ from $\mathbf{E}'=-\nabla V$, we would have to have $\nabla V=-\mathbf{v}\times\mathbf{B}^{\mathrm{m}\neq 0}$. But this choice is impossible because $\nabla\times\nabla V=\mathbf{0}$ whereas $\nabla\times(\mathbf{v}\times\mathbf{B}^{\mathrm{m}\neq 0})\neq\mathbf{0}$ by Eq. (8). The rotating nonaxisymmetric field does not, and cannot, produce a Lorentz force analogous to that produced by the axisymmetric field [7].

## II. CONDITIONS REQUIRED FOR VOLTAGE GENERATION

Because the m $\neq 0$ components of Earth's field produce no net force on charges rotating with Earth, they cannot be used to drive an electric current. However, the axisymmetric m $= 0$ components of Earth's total field (henceforth simply designated $\mathbf{B}$ for notational simplicity) do lead to a $q\mathbf{v} \times \mathbf{B}$ force, so we examine the results of this force further here. In any conductor carried by Earth's rotation, the effect of this force is to rapidly redistribute electrons [27], until the resulting electrostatic field $\mathbf{E} = -\nabla V$ perfectly cancels the driving force:

$$\mathbf{v} \times \mathbf{B} = \nabla V. \tag{12}$$

(Note $\nabla \times (\mathbf{v} \times \mathbf{B}^{\mathrm{m}=0}) = \mathbf{0}$ so Eq. (12) may be satisfied, unlike the case for $\mathbf{B}^{m\neq 0}$.) The classical charge relaxation timescale for this electron redistribution to occur is $\epsilon_0/\sigma \approx 10^{-11}$ $(1\ \mathrm{S\,m^{-1}}/\sigma)$ s [28], where $\epsilon_0$ is vacuum permittivity, $\sigma$ electrical conductivity, and for reference Earth's seawater has $\sigma = 3.3\ \mathrm{S\,m^{-1}}$ [29]. Equation (12) has been used to estimate depth- and latitude-dependent volume charge densities due to electron redistribution within the Earth itself as high as $\sim 1\ e^-\ \mathrm{m^{-3}}$ [27].

The extremely rapid and continuously ongoing field cancellation within any conductor appears to make it impossible to use the $\mathbf{v} \times \mathbf{B}$ force to generate electricity even for the m $= 0$ components of Earth's field. However, Eq. (12) hides an implicit assumption that can be violated [2]. Since $\nabla \times \nabla V = \mathbf{0}$ always, Eq. (12) cannot be satisfied within a magnetically permeable object with a topology chosen to ensure

$$\nabla \times (\mathbf{v} \times \mathbf{B}) \neq \mathbf{0} \tag{13}$$

within the object. The inequality of Eq. (13) holds, for example, in the interior region $a < \rho < b$ of a magnetically permeable conducting cylindrical shell of inner and outer radii $a$ and $b$ (with $\rho$ the radial coordinate in a cylindrical coordinate system centered in the shell), and long axis perpendicular to $\mathbf{v}$ and $\mathbf{B}$ (see Fig. 1). In such an object, by Eqs. (12) and (13) electrons cannot arrange themselves to cancel the $q\mathbf{v} \times \mathbf{B}$ force experienced as the shell is carried by Earth's rotation through Earth's $\mathbf{B}^{\mathrm{m}=0}$ field. Therefore, there are unbalanced forces, so there is a possibility of driving a continuous current [2].

The electromotive force (emf) around a path $C$ within the conducting shell is given by Eq. (3). But because of the $-\partial\mathbf{B}/\partial t$ term in Eq. (3), Eq. (13) does not guarantee a nonzero emf. Determining whether currents actually flow requires an examination of the integrand of Eq. (3) and the conditions under which it will be nonzero.

Any moving conductor that obeys Ohm's law satisfies [27,28]

$$\mathbf{E} + \mathbf{v} \times \mathbf{B} = \mathbf{J}/\sigma = \eta \nabla \times \mathbf{B}, \tag{14}$$

where $\mathbf{J}$ is the current density and magnetic diffusivity $\eta = (\sigma\mu)^{-1}$. Taking the curl of Eq. (14) gives the advection-diffusion equation for $\mathbf{B}$ within that conductor,

$$-\partial\mathbf{B}/\partial t + \nabla \times (\mathbf{v} \times \mathbf{B}) = -\eta \nabla^2 \mathbf{B} \tag{15}$$

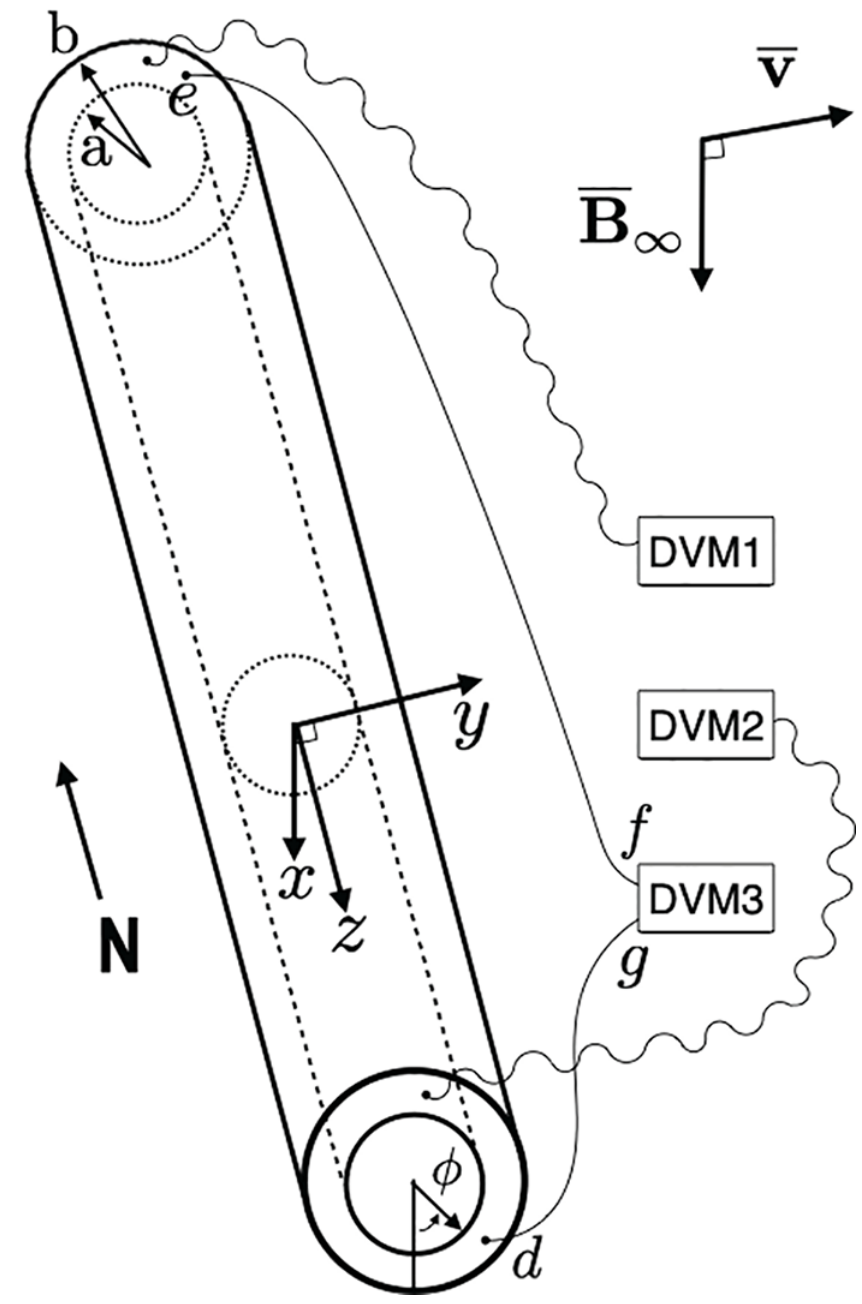

FIG. 1. Experimental configuration. Sketch shows orientation of low-$R_m$ MnZn ferrite cylindrical shell, with voltage (or current) electrodes and thermocouples attached. The long axis of the shell (along the $\hat{\mathbf{z}}$ coordinate) is orthogonal both to Earth's m $= 0$ magnetic flux density $\mathbf{B}_\infty = B_\infty\hat{\mathbf{x}}$ and linear velocity of rotation $\mathbf{v} = v\hat{\mathbf{y}}$. DVM1 and DVM2 connect to thermocouples and measure temperatures at opposite ends of the shell. DVM3 measures emf (across points $d$ and $e$) or electric current (around the circuit $defgd$) at experimenter's choice; point $f$ corresponds to its positive terminal. With point $e$ to the north, the system is at an orientation of 0°; when rotated so $e$ points east, the system is at 90°, and so on for 180° and 270°. These define the orientations referred to in Figs. 2–6.

for constant $\eta$. The left-hand side of Eq. (15) integrated over the relevant surface $S$ is evidently the emf around the corresponding path $C$ in Eq. (3). The right-hand side of Eq. (15), and so the emf in Eq. (3) must be negligible if $C$ lies within a conductor with magnetic Reynolds number $R_m = \sigma\mu v\xi \gg 1$, where $\xi$ is the length scale over which $\mathbf{B}$ changes appreciably within the conductor.

In summary, a nonzero emf is possible provided we choose a magnetically permeable conductor that satisfies two requirements [2]: a topology (such as a cylindrical shell) that satisfies Eq. (13); and a conductivity, permeability, and scale that yields

$$R_m = \sigma\mu v\xi < 1. \tag{16}$$

It is well-known that a magnetically permeable cylindrical shell oriented as in Fig. 1 perpendicular to the field $\mathbf{B}_\infty$ will distort that field near and within the shell. The resulting field for a shell that has velocity $\mathbf{v} = \mathbf{0}$ when there is no electric field present must satisfy, by Eq. (14),

$$\nabla \times \mathbf{B} = \mathbf{0}. \tag{17}$$

This in turn means that $\mathbf{B}$ can be written as the gradient of a magnetic potential $W$, $\mathbf{B} = -\nabla W$, so that $\nabla \cdot \mathbf{B} = 0$ yields a Laplace equation for $W$ that may be solved with boundary conditions on $\mathbf{B}$ and $\mathbf{H}$ for the shell to give the resulting field within and exterior to the shell [30,31]. We will call this field $\mathbf{B_0}$ to indicate that it is the $\mathbf{v} = \mathbf{0}$ solution. In this static situation, current flow within the shell will clearly not take place. But if a conducting magnetically permeable cylindrical shell has $\mathbf{v} \neq \mathbf{0}$, Eq. (14) instead yields not Eq. (17) but $\nabla \times \mathbf{B} = \eta^{-1}(\mathbf{E} + \mathbf{v} \times \mathbf{B})$, which cannot in general be satisfied by the choice $\mathbf{B} = -\nabla W$, and the familiar solution $\mathbf{B_0}$ is not correct. We must, instead, solve Eq. (15) in full. Shifting to the frame $K'$, in which Eq. (14) becomes $\mathbf{E}' = \eta\nabla \times \mathbf{B}'$, does not change this conclusion as it merely introduces an electric field $\mathbf{E}' = \mathbf{v} \times \mathbf{B}$.

If the permeable cylindrical shell has $R_m \gg 1$, then the distorted magnetic flux density within the shell interacts with charged particles $q$ within the conductor analogously to the way the m $\neq 0$ components of Earth's magnetic field do and $\mathbf{E} = -\mathbf{v} \times \mathbf{B}$, or very nearly so. Then $\mathbf{B_0}$ once again becomes the correct solution, at least to high precision. We will therefore focus on shells satisfying $R_m \leqslant 1$.

Equation (15) may be solved analytically when the shell is oriented as in Fig. 1 along geographic south-to-north, orthogonal to both $\mathbf{B}_\infty = B_\infty\hat{\mathbf{x}}$, the m = 0 part of Earth's magnetic flux density far from the shell, and to $\mathbf{v} = v\hat{\mathbf{y}}$, Earth's rotation velocity at the shell's location [2]. The $x$ component of that solution is [2]

$$\begin{aligned} B_{sx}(a < \rho < b) = \beta_1 - \beta_2(a/\rho)^2 e^{ky}\{[k\rho\cos 2\phi \\ + (k\rho)^2 \sin\phi]K_1(k\rho) \\ - (k\rho)^2\sin^2\phi K_0(k\rho)\}, \end{aligned} \tag{18}$$

where $y = \rho\sin\phi$, $K_0(k\rho)$ and $K_1(k\rho)$ are modified Bessel functions of the second kind of order 0 and 1, $2kb = \mu\sigma vb = R_m$, $\beta_1 = 2B_\infty\mu_r(\mu_r+1)[(\mu_r+1)^2 - (a/b)^2(\mu_r-1)^2]^{-1}$, and

$$\beta_2 = 2B_\infty\mu_r(\mu_r-1)[(\mu_r+1)^2 - (a/b)^2(\mu_r-1)^2]^{-1}. \tag{19}$$

Relative permeability $\mu_r$ is defined by $\mu = \mu_r\mu_0$, with vacuum permeability $\mu_0 = 1.257 \times 10^{-6}$ N A$^{-2}$.

The condition $R_m < 1$ corresponds to $k\rho < 1/2$ in Eq. (18), which may then be expanded by taking series expansions in $k\rho$ (corresponding to powers of $R_m$) for $e^{k\rho}$, $K_0(k\rho)$, and $K_1(k\rho)$ (see Appendix A or [2]). We find

$$\mathbf{B_s} = \mathbf{B_0} + \Sigma_{n=1}^{N}\mathbf{B_n} + O(R_m)^{N+1}, \tag{20}$$

where $\mathbf{B_0}$ is the usual magnetic flux density in $a < \rho < b$ for a magnetically permeable shell when $\mathbf{v} = \mathbf{0}$, and $\mathbf{B_n}$ ($n = 1, 2, 3...N$) is an $n$th-order in $R_m$ perturbation to $\mathbf{B_0}$ that enters when $\mathbf{v} \neq \mathbf{0}$.

### III. VOLTAGE GENERATION

The emf along a path $\bar{C}$ between the points $d$ and $e$ within the cylindrical shell is, by Eqs. (3) and (14),

$$\varepsilon = \eta\int_{\bar{C}}(\nabla \times \mathbf{B_s}) \cdot \mathbf{dl}, \tag{21}$$

where $\mathbf{B_s}$ is the magnetic flux density within the shell, i.e., for $\rho$ such that $a < \rho < b$. We previously showed [2] that there exist closed paths within the cylindrical shell around which emf $\neq 0$ to $O(R_m)$. We also noted, however, that it is unclear as a practical matter that attaching electrodes to the exterior of the shell would enable one to select or access any particular path, since the shell is a continuous homogeneous conductor and one expects current to flow over many possible interior paths. Therefore, in experimental practice we simply attach digital multimeter voltage or current leads to the ends of the shell, as depicted in Fig. 1, giving a path from $d$ to $e$ through the shell, to $f$ at the positive terminal of DVM3, through DVM3 to $g$, DVM3's negative terminal, then back to $d$.

This approach has the merit of simplicity: DVM3 connects to the ends of the cylindrical shell with clip leads, with the top of each clip at $\rho = b$ and the bottom at $\rho = a$. Given the configuration shown in Fig. 1, one can choose whether to measure voltage or current simply by turning the dial on the multimeter between μV and μA settings and appropriately changing the input jacks. This configuration in effect treats the cylindrical shell as a lumped circuit element. Equation (21) then manifests a key prediction: If the entire system in Fig. 1 is rotated by 180° about the $x$ axis, then the measured emf (subsequent to the decay of any eddy currents caused by the rotation) should change sign, but be of equal magnitude. What voltage magnitude do we predict?

The circuit segment in Fig. 1 running from $e$ to $d$ through DVM3 makes no contribution: any nonzero $\mathbf{v} \times \mathbf{B}$ in these segments causes an electron displacement that cancels that $\mathbf{v} \times \mathbf{B}$. We predict emf generation in the shell from $d$ to $e$ by evaluating Eq. (21), following [2] but extending that calculation to higher orders in $R_m$, and averaging over $\phi$ from 0 to $2\pi$ and $\rho$ from $a$ to $b$. Details of the calculation described below are given Appendix A.

For the configuration in Fig. 1 [2],

$$\eta\nabla \times B_s = -\nabla V - \eta\nabla^2\mathbf{A_s} = -\nabla V - vB_{sx}\hat{\mathbf{z}}, \tag{22}$$

where $V$ is the electric potential, $\mathbf{A_s} = A_z\hat{\mathbf{z}}$ is the magnetic vector potential within the shell, and $B_{sx}$ is just the $x$ component of $\mathbf{B_s}$. Equation (22) and the derivation below treat the shell as infinitely long. For an actual finite-length shell, there are end effects that fall off exponentially with distance $x$ into the interior like $\exp(-3.83z/a)$ for interior radius $a$ [32]. Therefore, in our experiments we use a cylinder satisfying $L \gg 2a$.

We substitute Eq. (22), using the expansion Eq. (20), into Eq. (21) and average over $\phi$ and $\rho$. We find that $\nabla V + vB_{0x}\hat{\mathbf{z}}$, $B_{1x}\hat{\mathbf{z}}$, and $B_{3x}\hat{\mathbf{z}}$ in Eq. (22) averaged over $\phi$ from 0 to $2\pi$ each give zero. However, the second-order term is

$$\begin{aligned} B_{2x}(a < \rho < b) = -(\beta_2/8)R_m^2(a/b)^2[\ln(k\rho/2) \\ + 4\sin^2\phi - 2\sin^4\phi + \gamma - 1/2], \end{aligned} \tag{23}$$

which does not average to zero over $\phi$ and $\rho$:

$$\begin{aligned} \langle B_{2x}\rangle_{\rho,\phi} &= \frac{1}{\pi(b^2-a^2)}\int_0^{2\pi}\int_a^b B_{2x}\rho d\rho d\phi \\ &= \frac{\beta_2}{8}R_m^2\left(\frac{a}{b}\right)^2\left[\ln\left(\frac{4}{R_m}\right) + \frac{a^2}{b^2-a^2}\ln\left(\frac{a}{b}\right) \right. \\ &\quad \left. -\gamma - \frac{1}{4}\right]. \end{aligned} \tag{24}$$

Then from Eqs. (21) and (22) we find

$$\langle \varepsilon \rangle_{\rho,\phi} = v(\beta_2/8) l R_m^2 (a/b)^2 [\ln(R_m/4) - (a/b)^2 [1 - (a/b)^2]^{-1} \times \ln(a/b) + \gamma + 1/4] + O(R_m)^4, \tag{25}$$

where $\gamma = 0.5772...$ is Euler's constant, and $l$ is the distance between points $d$ and $e$. By Eqs. (25) and (19), $\langle \varepsilon \rangle_{\rho,\phi} = 0$ if $a = 0$ (as must be so since $\nabla \times (\mathbf{v} \times \mathbf{B}) = \mathbf{0}$ for a solid cylinder), or if $\mu_r = 1$. At Princeton's location $B_\infty = 45\,\mu\text{T}$ (obtained by summing the dipole, quadrupole, and octupole components of Earth's m = 0 field [2]) pointed downward into Earth's surface at an angle (from the horizontal when facing the north geographic pole) of 57.5°, and the direction to the geographic pole is currently 12.6° east of magnetic north [33]. For $\mu_r \gg 1$, $\beta_2 \approx 2B_\infty/[1 - (a/b)^2] = 143\,\mu\text{T}$.

## IV. MATERIALS AND METHODS

To test the prediction that electricity can be produced by a system that satisfies both Eqs. (13) and (16), we use a cylindrical shell made of MnZn ferrite, a soft magnetic material with conductivity about that of seawater [34]. This cylindrical shell and an otherwise identical solid cylinder used as a control were produced to our order by National Magnetics Group Inc. (see Appendix B). Our shell of M100 MnZn ferrite has length $L = 29.9$ cm and inner and outer radii satisfying $a/b = 0.61$ with $b = 1.0$ cm. We determine the temperature-dependent relative permeability for M100 MnZn ferrite from its data sheet values [35] as described in Appendix B. We have $\mu_r = 9,500 \pm 2,850$ for the temperatures at which we conducted our experiments. Ambient environmental conditions in our laboratory during our experiments are described in Appendix C. We determine the shell's conductivity using voltage and current measurements as described in Appendix B; we find $\sigma = 2.07 \pm 0.22\ \text{S m}^{-1}$. We take the shell's characteristic diffusion length scale to be $\xi = b$, so

$$R_m = \sigma \mu_r \mu_0 v b. \tag{26}$$

At Princeton's latitude our shell has $R_m \approx 0.088 \pm 0.028$, satisfying Eq. (16).

The cylinder and cylindrical shell were mounted on a plexiglass turntable on a wood base (that is, no conducting or magnetizable materials were used). The rotation axis of the turntable coincided with the origin of an underlying polar coordinate system. The turntable and underlying base were oriented and tilted to be perpendicular to $B_\infty$, as depicted in Fig. 1. Voltages and temperatures were recorded using three battery-operated Gossen Metrawatt Metrahit 30M digital multimeters (labeled in Fig. 1 and elsewhere as DVM3 for the voltage (or currrent) measuring device and DVM1 and DVM2 for the temperature measuring devices). The Metrahit 30M provides voltage and temperature measurement precisions of 0.1 µV and 0.01 °C.

Each experimental run for a particular orientation of the cylinder or cylindrical shell was begun by rotating the shell to its appropriate position and then allowing the system to sit overnight. Voltmeter leads were fixed to the plexiglass turntable, and were rotated together with the shell or cylinder. The multimeters also moved with the rotation. Therefore, there was no circuit-topology change under rotation.

Measurements were typically begun in the morning following a period of at least 8 h subsequent to the shell's rotation into position. Data were recorded on all three multimeters every 10 s for 5–10 h periods. The leads were then disconnected from the multimeters and data downloaded for analysis. Each step in this process is described in greater detail in Appendix D.

## V. PREDICTIONS FOR A SIMPLE LABORATORY SYSTEM

Equation (25) predicts the direct-current emf that DVM3 in Fig. 1 should measure, apart from a thermoelectric contribution (discussed below). For M100 material with $b = 1.0$ cm, $a/b = 0.61$, and $l = 27.9$ cm ($l < L$ because the DVM3 electrodes each extend 1.0 cm in from the ends of the shell), and propagating the uncertainty due to $\mu_r$ and $\sigma$, we predict $\langle \varepsilon \rangle_{\rho,\phi} \approx -13.7 \pm 7.2\,\mu\text{V}$ for the configuration depicted in Fig. 1, the orientation we define to lie at 0°. A rotation by 180° about the $x$ axis should yield the same emf but of opposite sign.

By symmetry, we expect the shell in Fig. 1, when rotated to 90° or 270°, to give emf = 0 µV. This can also be predicted analytically. For a shell aligned at 270°, for example, we have $\mathbf{v} = v\hat{\mathbf{z}}$ and $\mathbf{B} = (B_x, B_y, 0)$ in the shell, with $B_z = 0$ as long as we ignore end effects. Explicit calculation then shows $\nabla \times (\mathbf{v} \times \mathbf{B}) = \mathbf{0}$, given $\nabla \cdot \mathbf{B} = 0$ and provided no component of $\mathbf{B}$ varies with $z$ (again correct apart from end effects [2,32]). But when $\nabla \times (\mathbf{v} \times \mathbf{B}) = \mathbf{0}$, electron redistribution cancels the $q\mathbf{v} \times \mathbf{B}$ force, giving emf = 0.

We have therefore so far identified three testable predictions for the M100 cylindrical shell depicted in Fig. 1: (1) when the cylindrical shell is oriented at 0°, the system should generates a particular DC emf that lies in the range $-13.7 \pm 7.2\,\mu\text{V}$; (2) this emf should reverse sign upon rotation about the $x$ axis to 180°; and (3) emf should be 0 when the system is rotated to 90° or 270°. We can also propose two further tests: (4) for an otherwise identical solid ($a = 0$) M100 cylinder, Eq. (25) predicts emf = 0 for all orientations; and (5) by Eqs. (3) and (14), a cylindrical shell with $R_m \gg 1$ should produce zero voltage at all orientations.

## VI. EXPERIMENTAL RESULTS

### A. Controlling for the Seebeck voltage

Testing these predictions requires controlling for potentially confounding Seebeck voltages. A temperature gradient $\Delta T$ along the cylindrical shell between the two DVM3 electrodes will lead to a Seebeck voltage

$$\varepsilon_S = -S\Delta T, \tag{27}$$

where $S$ is the Seebeck coefficient. For many ordinary conducting metals, $S \sim 1 - 10\,\mu\text{V K}^{-1}$; for semiconductors the sign of $S$ can be positive or negative depending on the dopant. MnZn ferrites have the general formula $Mn_xZn_{1-x}Fe_2O_4$, with electrical properties varying greatly with $x$ [34,36]. Unfortunately, MnZn ferrites have especially large Seebeck coefficients, with published $S$ values for five different MnZn ferrites in the range $\pm(500 - 800)\,\mu\text{V K}^{-1}$ [36]. As described below, our measurements yield $S = -417\,\mu\text{V K}^{-1}$ for M100 material. The Cu DVM3 leads have such a low comparative

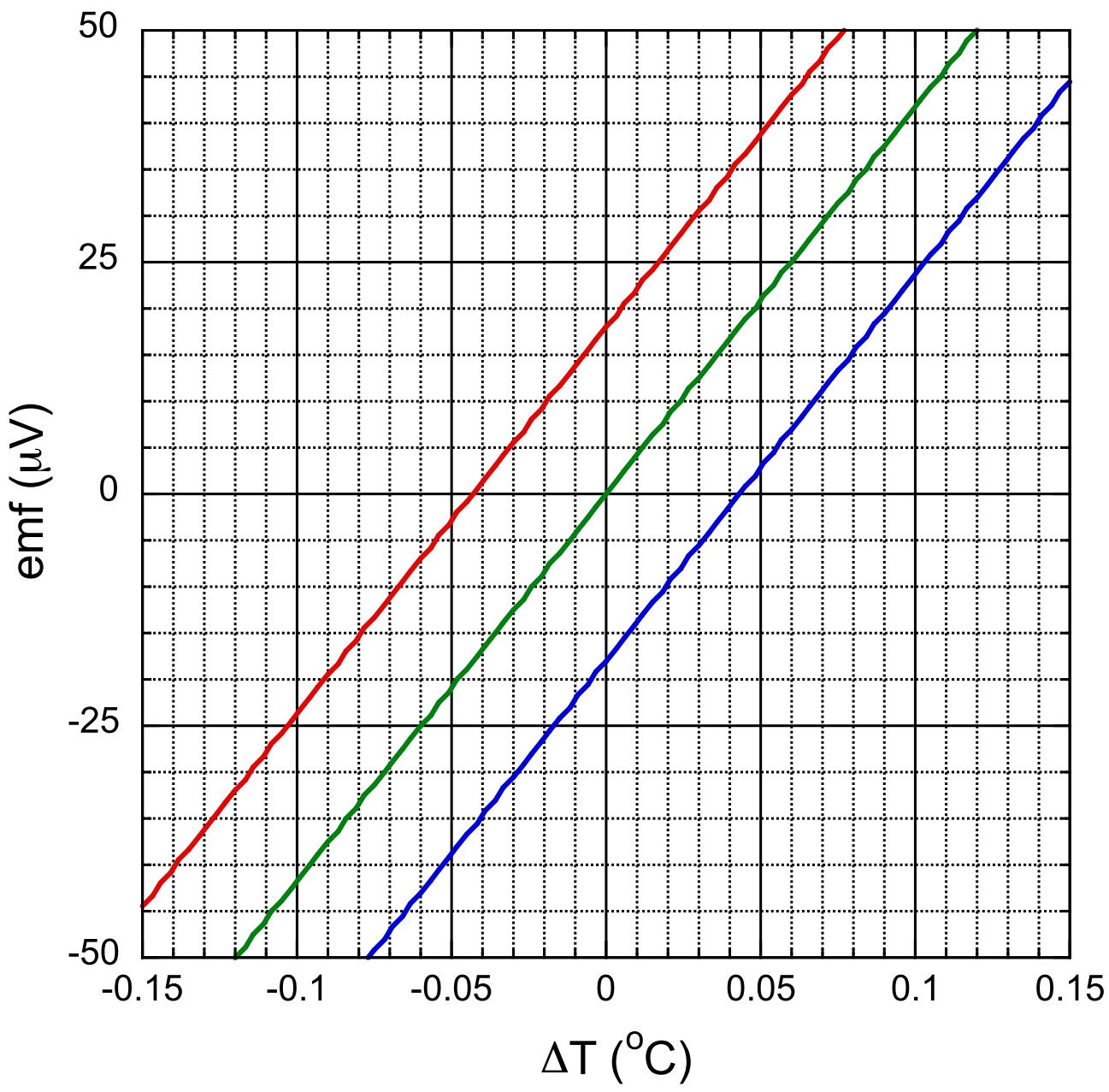

FIG. 2. Predicted emf behavior for a low-$R_m$ magnetically permeable cylindrical shell (Fig. 1). Voltage varies with $\Delta T$ along the shell due to the Seebeck effect [Eq. (27)], illustrated here by the green line for a Seebeek coefficient $S = -417\,\mu\text{V K}^{-1}$. Absent the effect predicted in this paper, the shell would generate voltages lying on this line for all orientations. But if the predicted effect exists and generates, say, a voltage of $-18\,\mu\text{V}$ [Eq. (25)], voltages would still lie along the solid green line at 90° and 270°orientations, but would lie along lines offset from the green line by $\langle\varepsilon\rangle_{\rho,\phi} = +18\ \mu\text{V}$ at 180° (red line) and by $\langle\varepsilon\rangle_{\rho,\phi} = -18\ \mu\text{V}$ at 0° (blue line). By contrast, an otherwise identical $a = 0$ (solid) $R_m < 1$ cylinder should produce voltages that lie along the green line at all orientations. Results for experimental tests of these predictions are shown in Figs. 3 and 4.

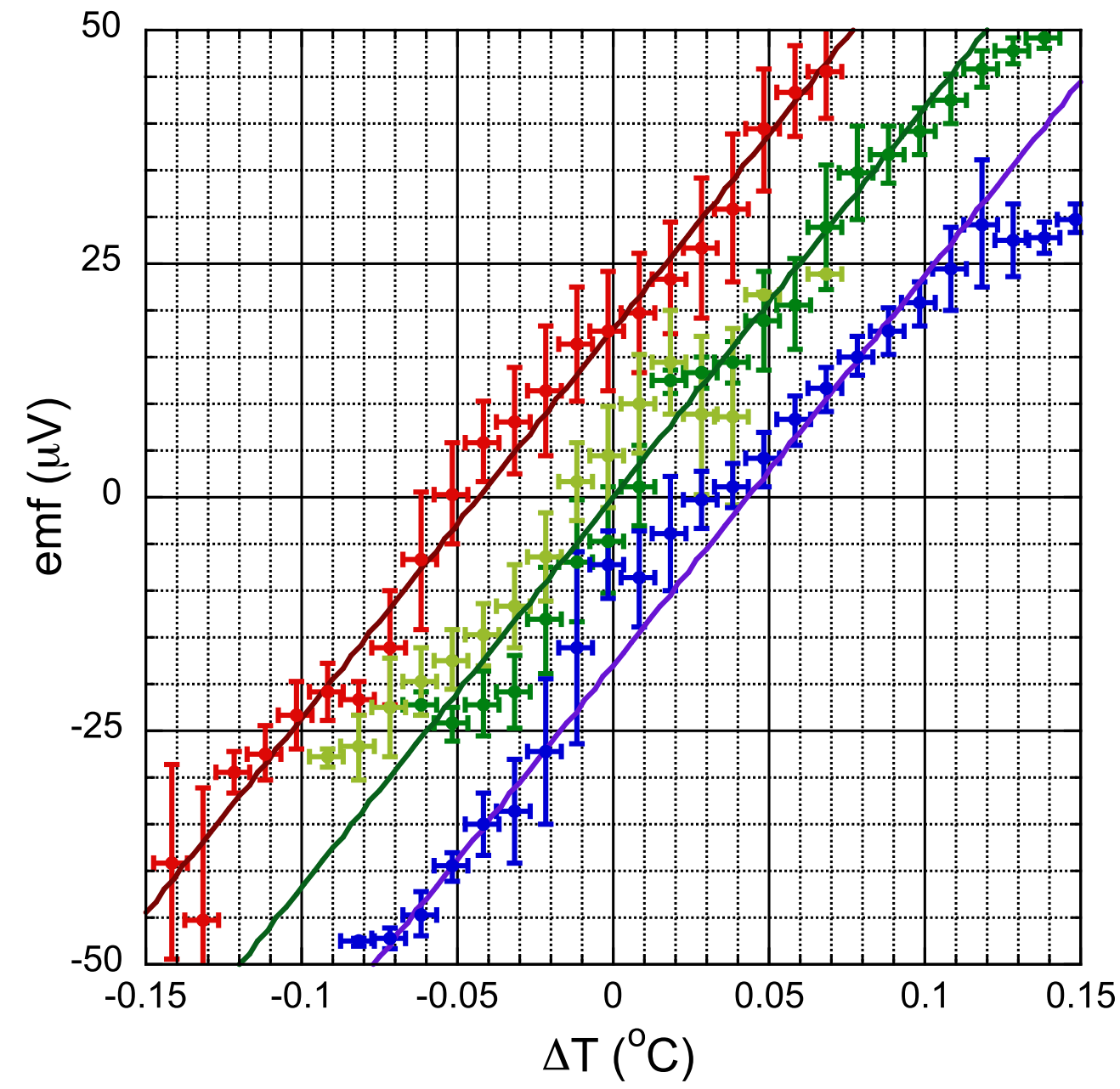

FIG. 3. Measured emf behavior (points with error bars) for the $a/b = 0.61$ M100 cylindrical shell, with the predictions of Fig. 2 overlying the data. Voltages are plotted against $\Delta T$ for four orientations of the shell depicted in Fig. 1: (1) 0° (blue); (2) 90° (dark green); (3) 180° (red); and (4) 270° (light green). Error bars are one sigma (plus pythagorean sum with small scale error; see Appendix D) for measurements at a given $\Delta T$ value for that orientation. As predicted in Fig. 2, orientations (2) and (4) show the effects of Seebeck voltages $\varepsilon_S$ only [Eq. (27)], whereas orientations (1) and (3) show predicted $\Delta T$-independent additional voltages $\langle\varepsilon\rangle_{\rho,\phi}$ [Eq. (25)] of appropriate sign and magnitude.

Seebeck coefficient, $S = 1.9\,\mu\text{V K}^{-1}$ at room temperature [37], that we ignore this small additional effect.

Ambient vertical and horizontal temperature gradients in our laboratory can reach $\sim 1\,^\circ\text{C m}^{-1}$ (see Appendix C), leading to values for $\Delta T$ along the cylindrical shell as large as $0.3\,^\circ\text{C}$. By Eq. (27), this could result in Seebeck voltages as high as $\varepsilon_S \sim \pm 120\,\mu\text{V}$, obscuring the $-14 \pm 7\,\mu\text{V}$ signal we expect from our effect. Moreover, a persistent ambient $\Delta T$ could mimic the predicted sign reversal in the emf generated by our effect when the system is rotated.

It is therefore essential that we control for $\varepsilon_S$ as a function of $\Delta T$. As shown in Fig. 1, simultaneously with recording emf (or current) values by DVM3, we also record temperatures $T_1$ and $T_2$ measured by DVM1 and DVM2, respectively, using thermocouples in thermal contact with opposite ends of the shell. DVM3 records an emf $= \langle\varepsilon\rangle_{\rho,\phi} + \varepsilon_S$, and as illustrated in Fig. 2 we can plot this against $\Delta T = T_1 - T_2$. When $\langle\varepsilon\rangle_{\rho,\phi} = 0$, as should be the case for both the $a = 0$ cylinder at any orientation, and for the $a/b = 0.61$ shell in a 90° or 270° orientation, our data should fall along a line (the green solid line in Fig. 2) that passes through the origin: emf $= 0$ when $\Delta T = 0$. If the effect predicted in this paper did not exist, then the $a/b = 0.61$ shell should also generate voltages lying on this green line at any orientation. But if the effect exists, then distinct parallel lines offset from the green line should result for 180° (red line) and 0° (blue line) due to emf generation at these orientations.

In Appendix C we describe additional laboratory measurements to exclude the possibility that the DC voltages and currents that we generate could somehow be due to 60 Hz or RF background.

### B. Testing five predictions

Experimental results for the $a/b = 0.61$ cylindrical shell are shown in Fig. 3 and compared with the predictions of Fig. 2. As predicted in Fig. 2, orientations of 90° and 270° show the effects of Seebeck voltages $\varepsilon_S$ only [Eq. (27)], whereas orientations of 0° and 180° show predicted additional voltages $\langle\varepsilon\rangle_{\rho,\phi}$ [Eq. (25)] of appropriate, and equal and opposite, magnitude. As also predicted by Eq. (25), the sign of the voltage displacement measured for the 0° orientation is negative, whereas the voltage displacement measured by the cylindrical shell when rotated to 180° is positive.

As discussed in greater detail in Appendix B, voltmeter leads were secured to the plexiglass turntable upon which the cylindrical shell was mounted, so that the topology of the multimeter-leads-shell circuit did not change under rotation. This was done to rule out any possible circuit-topology effects that, at least in certain $\partial \mathbf{B}/\partial t \neq \mathbf{0}$ circuits, can alter voltage magnitude and sign [38].

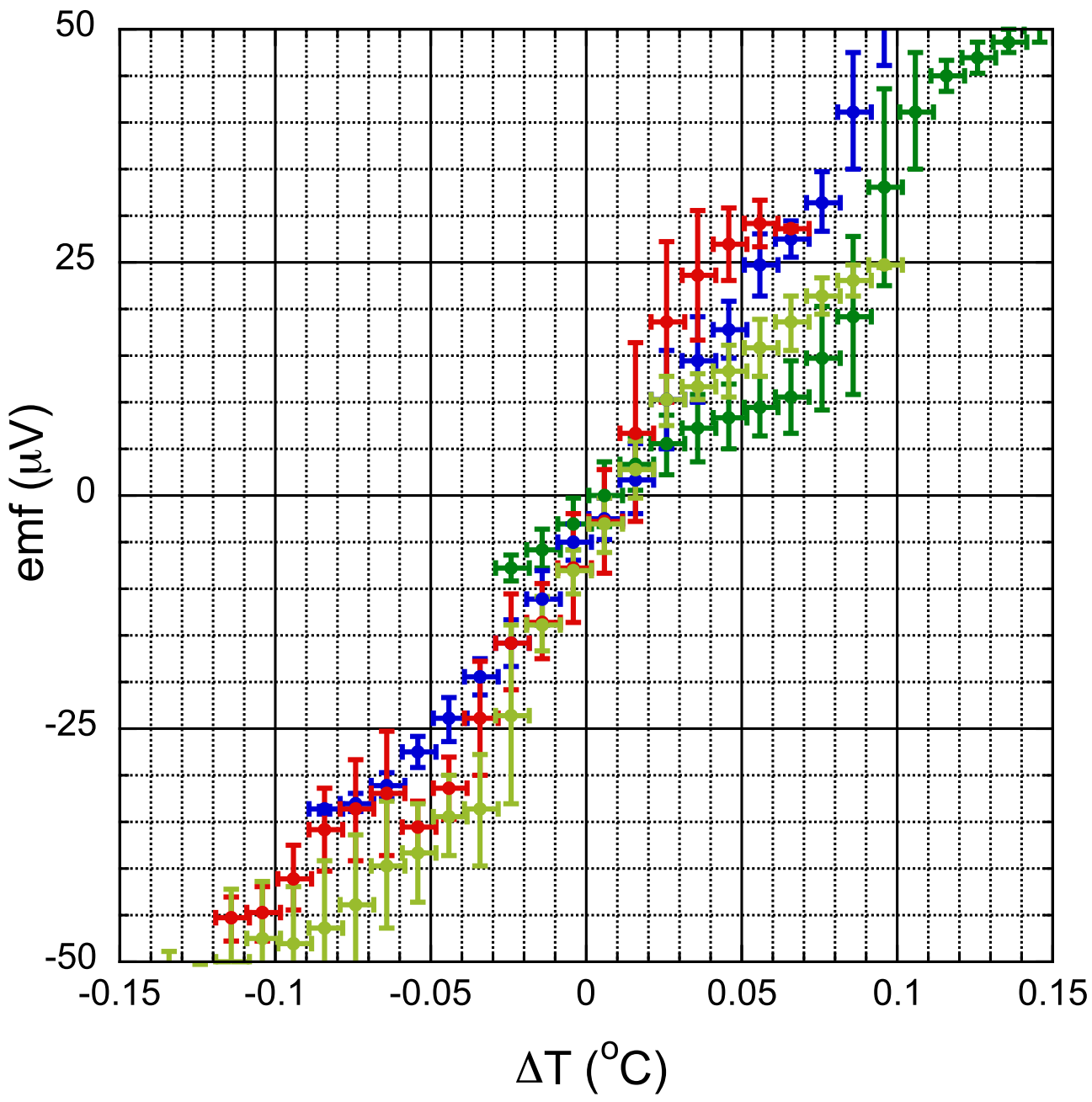

FIG. 4. Experimental results for the $a = 0$ solid M100 cylinder. Colors correspond to the same orientations as in Fig. 3. As predicted in the text, this solid cylinder, otherwise identical to the shell used in Fig. 3, yields $\langle \varepsilon \rangle_{\rho,\phi} = 0$ for all four orientations, so that the three separated lines of data predicted in Fig. 2 and found in Fig. 3 should collapse into a single line—as found here, albeit with scatter. The solid cylinder thus provides a successful negative control test for our results.

Results for the $a = 0$ control are shown in Fig. 4. By Eq. (25), this solid cylinder, otherwise identical to the shell used for Fig. 3, is predicted to generate $\langle \varepsilon \rangle_{\rho,\phi} = 0$ for all four orientations, so that the three separated lines of data predicted in Fig. 2 should collapse into a single line, as is found. It is clear from Figs. 3 and 4 that results for our system strongly corroborate our predicted effects.

In Appendix E, we plot in Fig. 6 the combined data of Fig. 4 for the $a = 0$ cylinder together with the combined 90° and 270° data of Fig. 3 for the $a/b = 0.61$ cylindrical shell, showing that these coincide. The Seebeck coefficient is given by regression over the former points, since the voltage for the $a = 0$ cylinder should be purely due to the Seebeck effect. Regression over these points in Fig. 6 gives $S = -417\,\mu\text{V K}^{-1}$. A regression over the data in Fig. 3 for the 180° orientation for the $a/b = 0.61$ shell gives emf $= 18.2\,\mu\text{V} + 419\,(\mu\text{V}/^\circ\text{C})\,\Delta T$ (R = 0.992), showing both the magnitude of $\langle \varepsilon \rangle_{\rho,\phi}$ and a Seebeck coefficient nearly identical to that found for the $a = 0$ cylinder. Regression over the data for emf at the orientations of 0° and 180° yields $|\langle \varepsilon \rangle_{\rho,\phi}| = 17.3 \pm 1.5\,\mu\text{V}$.

Our last prediction was that no effect should be observed for a magnetically permeable cylindrical shell having $R_m \gg 1$. We tested this using a MuMetal shell with $R_m \approx 10^6$, and measured emf $= 0.019 \pm 0.099\,\mu\text{V}$ at 0° orientation (averaging data over a 9.5 h run), confirming the prediction. (See Appendixes C and E for material and experimental details, respectively.)

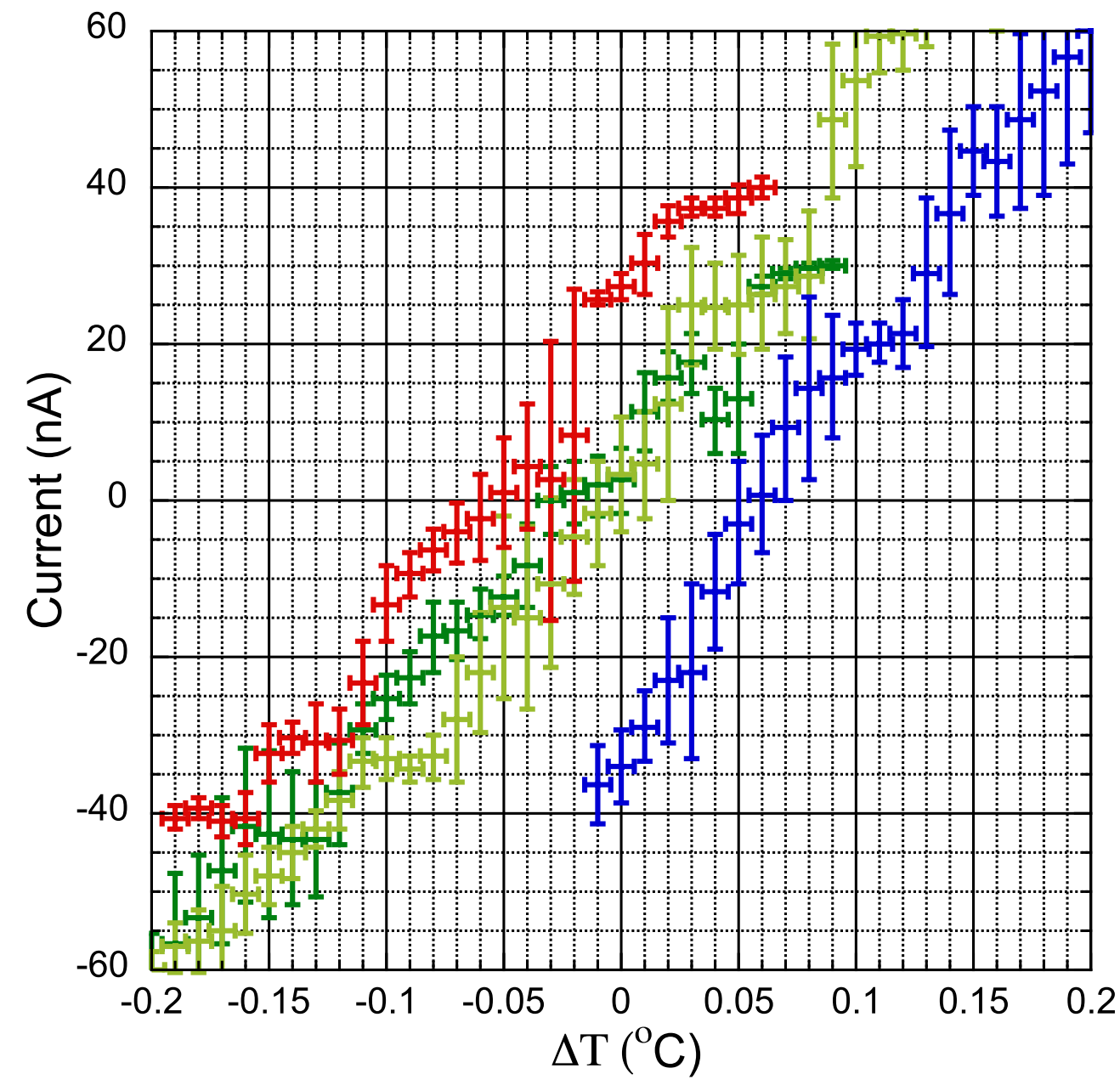

FIG. 5. Measured current behavior for the $a/b = 0.61$ M100 cylindrical shell. Currents are plotted against $\Delta T$ for four orientations of the shell depicted in Fig. 1: (1) 0° (blue); (2) 90° (dark green); (3) 180° (red); and (4) 270° (light green). Error bars are one sigma (plus pythagorean sum with small scale error; see Appendix D) for measurements at a given $\Delta T$ value for that orientation. The system exhibits analogous behavior regardless of whether the multimeter DVM3 in Fig. 1 is used as a voltmeter (giving the results in Fig. 3) to measure emf between points $d$ and $e$ in Fig. 1, or as an ammeter to measure current (as for this figure) around the circuit $defgd$ in Fig. 1.

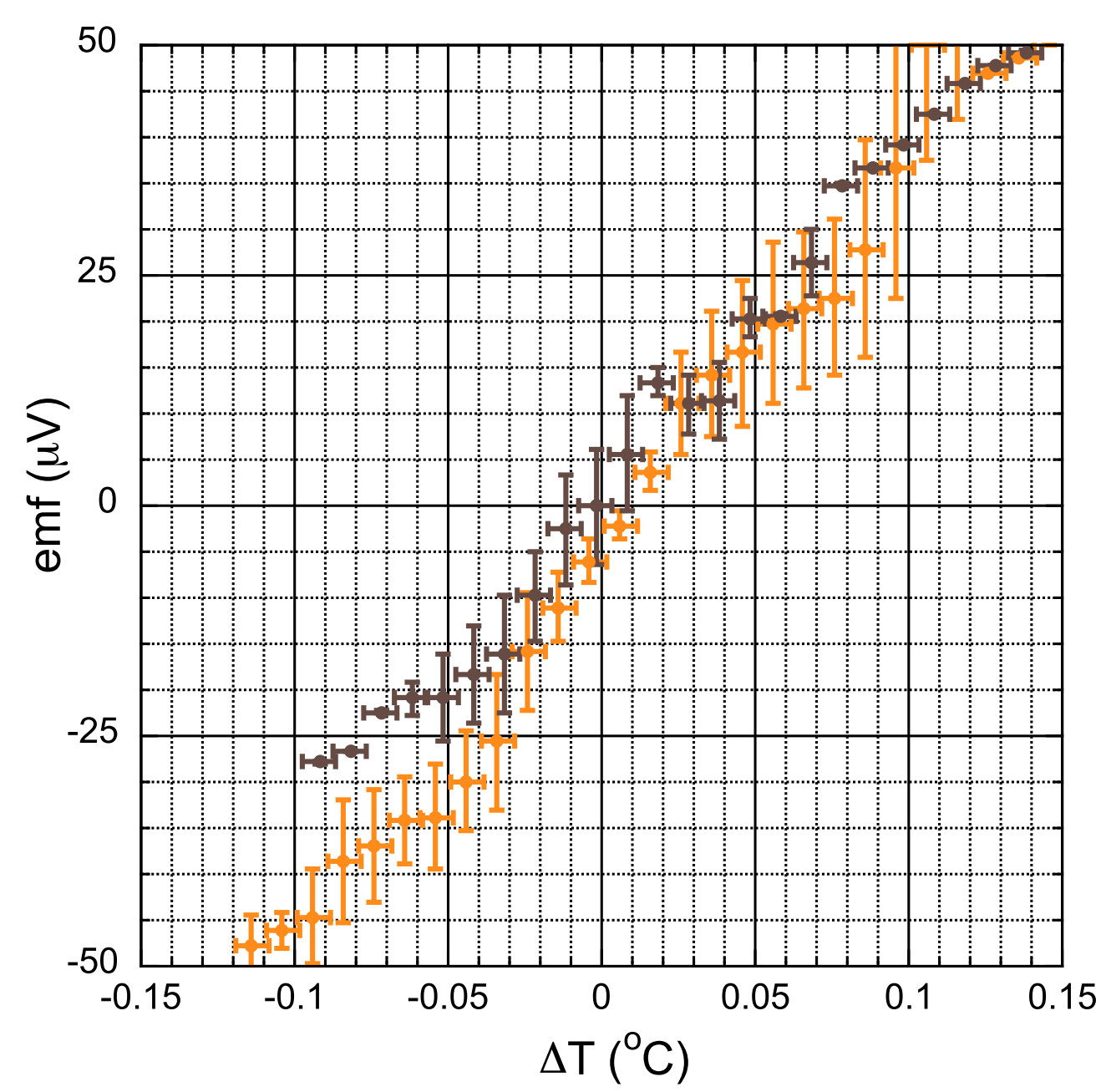

FIG. 6. Comparison of combined data from M100 solid cylinder ($a = 0$) at all four orientations in Fig. 4 (orange points) with combined data from M100 cylindrical shell ($a/b = 0.61$) at 90° and 270° orientations in Fig. 3 (brown points).

### C. Current measurements

Equation (25) predicts an emf that should be generated by our system and measurable by the multimeter DVM3 in voltmeter mode across points $d$ and $e$ in Fig. 1. It is also clear from Fig. 1 that DVM3 could instead be used in ammeter mode (see Appendix D) to measure current flow around the circuit $defgd$. We therefore conducted a few experiments in which we measured current (in nA) as a function of $\Delta T$. The results of these experiments are shown in Fig. 5, demonstrating that our system is generating a continuous DC current in addition to a continuous DC voltage—and therefore is generating DC power. Regression over the data for current at the orientations of 0° and 180° yields $|\langle i\rangle_{\rho,\phi}| = 25.4 \pm 1.5$ nA.

### D. Results reproduced at second location

The experiments providing the data underlying Figs. 3 and 4 were performed in an environmentally well-controlled underground windowless dark laboratory with low 60 Hz and RF backgrounds, as detailed in Appendix C. As a final check to test whether an unaccounted-for local effect could somehow be spoofing the entire array of predicted results obtained in that laboratory, we reproduced our experiments in a residential building 5.5 km east of our original experimental setting. This was a largely unregulated environment, in contrast to that of our primary laboratory. As described in Appendix F, the resulting data are noisy with correspondingly large error bars in comparison with the results obtained in our primary laboratory. Nevertheless, the data once again show the voltage magnitude and behavior under rotation predicted by our effect (Fig. 7), demonstrating that the observed effect is not due to an unidentified local influence in our primary laboratory.

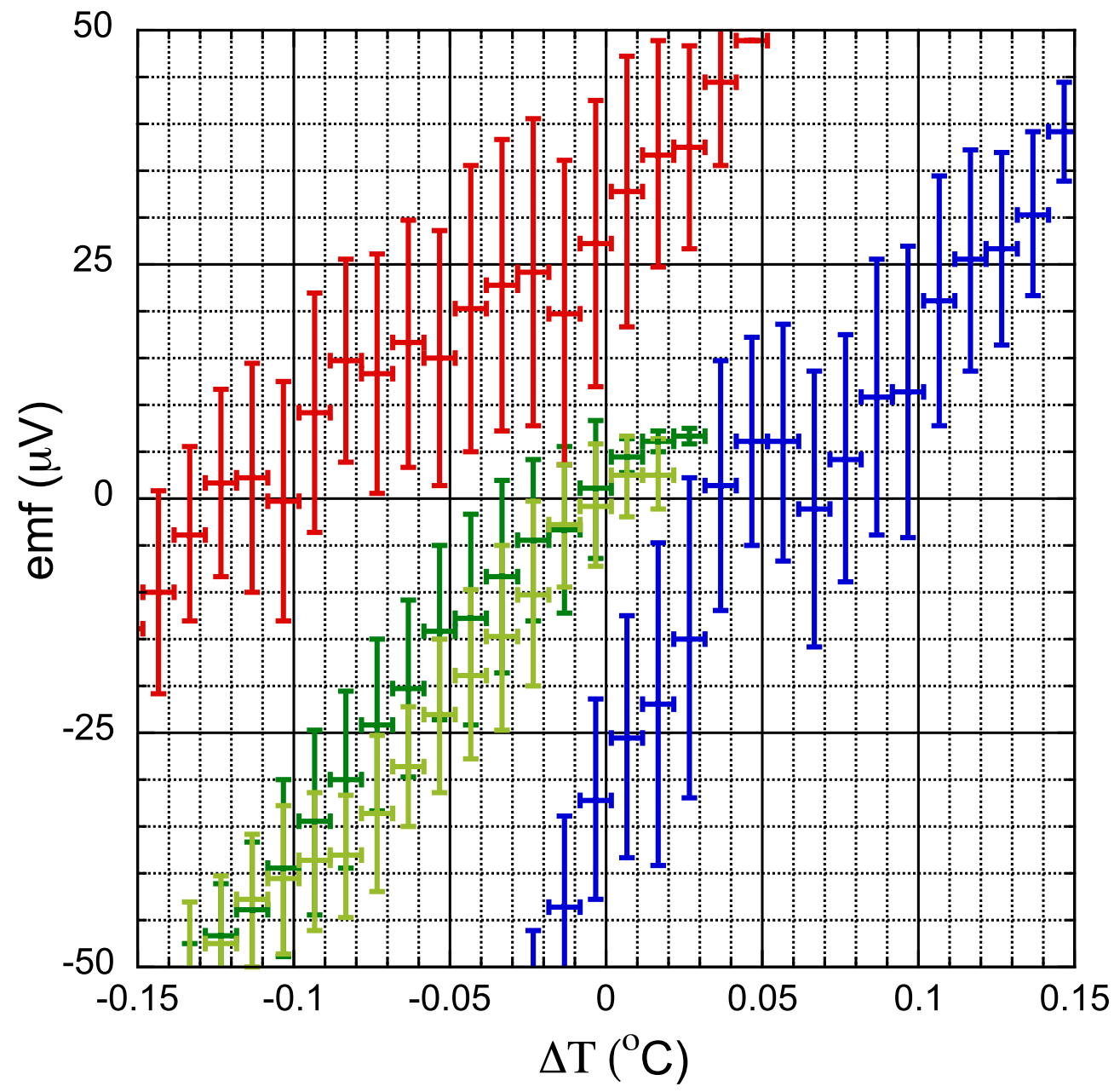

FIG. 7. Measured emf behavior for the $a/b = 0.61$ M100 cylindrical shell at Site B, an environmentally largely uncontrolled residential location 5.5 km east of our original laboratory. Voltages are plotted against $\Delta T$ for four orientations of the shell depicted in Fig. 1: (1) 0° (blue); (2) 90° (dark green); (3) 180° (red); and (4) 270° (light green). Error bars are one sigma (plus pythagorean sum with small scale error; see Appendix E) for measurements at a given $\Delta T$ value for that orientation. Measured voltages at Site B, while significantly noisier than those obtained in our laboratory, exhibit the same behavior as those taken in our laboratory, and as predicted for our effect.

## VII. PREVIOUS EXPERIMENTS AND OBJECTIONS

The effects predicted in Ref. [2] appear to be strongly corroborated by the experimental results presented here. But Veltkamp and Wijngaarden [39] presented results of experiments motivated by Ref. [2] that they argued gave null results and refute those predictions. However, none of their three experiments satisfied our criteria for voltage generation.

Veltkamp and Wijngaarden's experiment that most closely resembled the experiment that we proposed used a MnZn ferrite cylindrical shell with $b = 0.94$ cm, $a = 0.51$ cm, and $L = 2.86$ cm, and rotated this shell 180° not about the $x$ axis, as in our experiments, but about the cylinder's axis of symmetry, the $z$ axis (see Fig. 1 for these axis definitions). They measured an emf of amplitude 1.5 μV that tracked their rotations, consistent with the magnitude of the voltage for their system that they predicted were our effect real. But they argued, in light of the large Seebeck coefficients for MnZn ferrites, that their observations were likely due to a Seebeck voltage driven by a small $\Delta T$ experienced by the shell. Veltkamp and Wijngaarden [39] did not report temperature measurements during their experiments, but the underlying Master's Thesis by Veltkamp [40] describes ambient laboratory temperature measurements broadly consistent with a Seebeck voltage of the observed magnitude. It is unclear, however, how to disentangle the two effects potentially contributing to their voltage measurement in the absence of a control for the Seebeck effect.

But the predictions in Ref. [2] may well not hold for their system, even were it rotated about the $x$ axis, and even were there a control for the Seebeck effect. We required $L \gg 2a$ for the cylindrical shell in deriving our predictions. For our experimental system, $L = 29.9$ cm and $a = 0.6$ cm giving $L = 50a$, satisfying this requirement, whereas the system in Ref. [39] has $L = 2.8a$. Edge effects distort $\mathbf{B}$ within the shell away from the $\mathbf{B_0}$ that underlies our calculations [2,32], so one cannot expect that our predictions will hold within the system in Ref. [39].

Veltkamp and Wijngaarden also performed two other experiments that they presented as contradicting our predictions. The first of these involved creating a copper wire circuit in conductive contact with a cylindrical MnZn shell; the second was similar except with the Cu circuit insulated from the shell. However, these experiments are distinct from anything we proposed, so do not provide a test of our suggestion. In fact, we would expect these experiments to yield zero emf as found in Ref. [39]. We had speculated, at the end of our paper [2], that one might coat an underlying magnetically permeable shell with an "overlying shell" of a thin Cu layer to realize our effect. But we then stated that to achieve $R_m < 1$, given the high conductivity of Cu ($\sim 6 \times 10^7$ S m$^{-1}$), that apparatus would have to be microscopic in size, with $b < 5$ μm. Veltkamp and Wijngaarden [39] used a wire of diameter 0.4 mm in their experiments, giving $R_m \approx 10$ for their Cu circuit, so that we would expect the measured emf in these experiments to be zero, as they found.

Jeener [41] claimed to show that the integrand in Eq. (3) must always equal 0 for our system, so that no emf could be generated. We argued in our reply [7] that Jeener's conclusion was incorrect because his argument failed to recognize the distinction (reviewed above) between the physical effects of the $\mathbf{B}^{\mathrm{m}=0}$ and $\mathbf{B}^{\mathrm{m}\neq 0}$ terms upon a conductor corotating with Earth. In effect, Jeener treats all magnetic field components as behaving like the $\mathbf{B}^{\mathrm{m}\neq 0}$ components, with $\partial \mathbf{B}/\partial t$ always perfectly canceling $\nabla \times \mathbf{v} \times \mathbf{B}$.

It is natural to query energy and angular momentum conservation for our system. In Ref. [2] we showed by a Poynting vector ($\mathbf{S} = \mu^{-1}\mathbf{E} \times \mathbf{B}$) calculation that there is net power flowing into the cylindrical shell provided $\mathbf{v} \neq \mathbf{0}$ (or in a frame where $\mathbf{v} = \mathbf{0}$ but in which there is an equivalent electric field $\mathbf{E} = \mathbf{v} \times \mathbf{B}$). By explicit calculation, we showed that this power inflow equals the power lost from Earth's kinetic energy of rotation due to the $\mathbf{J} \times \mathbf{B}$ magnetic braking resulting from this device. (This is reminiscent of what takes place in a homopolar generator [42].) Therefore, energy is conserved: the energy driving currents within the shell ultimately derives from Earth's rotational kinetic energy, mediated by Earth's magnetic field. (Based upon the measured values of the emf and current, the configuration described here—merely an initial experimental demonstration system—taps into only a fraction of the available power. Raising this efficiency is a subject for subsequent investigation.) By contrast, for a solid cylinder with $a = 0$ there is no net Poynting energy inflow, and also no magnetic braking so no despinning of Earth.

A related issue is whether the slight despinning of Earth caused by the cylindrical shell is also consistent with angular momentum conservation. In electrodynamics energy and angular momentum reside in both the electromagnetic field and mechanical rotation, with the field angular momentum density equal to $\epsilon_0 \mu \mathbf{r} \times \mathbf{S}$ (where $\mathbf{r}$ is the usual radial component in a spherical coordinate system) [43]. Mechanical systems can then increase or decrease their angular momentum via the transfer of angular momentum with the electromagnetic field. We discussed this for our system in Ref. [2], but intuition may be provided by a thought experiment involving an initially nonrotating magnetized charged conducting spherical shell [43–45]. Angular momentum resides in the static $\mathbf{S}$ field of this nonrotating sphere ($\mathbf{E}$ is produced by the charge, $\mathbf{B}$ by the magnetization, and $\epsilon_0 \mu \mathbf{r} \times \mathbf{S} \neq \mathbf{0}$). The charge of the sphere is then drained to ground via the sphere's south pole. The sphere goes into rotation due to the $\mathbf{J} \times \mathbf{B}$ force during this draining, and it can be shown [43,44] that the angular momentum that had previously resided in the field (now zero because now $\mathbf{E} = \mathbf{0}$, since the sphere's charge has been drained) is identical to the mechanical angular momentum acquired by the newly rotating sphere.

We previously showed that even in an extreme scenario where our civilization somehow would obtain all its electrical energy from the effect described here, Earth's rotation would slow by $<1$ ms per decade [2]. By comparison, the length of Earth's day fluctuates by several ms per decade, likely due to interior mass redistributions and core-mantle coupling effects [46]. Earth's solid iron core rotates independently of the bulk Earth, changing from super-rotation to subrotation on decadal timescales [47]. Earth is also despinning due to exchange of angular momentum with the Moon, lengthening Earth's day currently by 2.5 ms per century [48].

## VIII. POSSIBLE SCALING TO HIGHER VOLTAGES

Results for our simple laboratory demonstration systems appear strongly to confirm the effects predicted by Eq. (25), as do the proposed relevant control experiments. The results have been confirmed at a second location in the same geomagnetic environment. The next step would be for an independent group to reproduce (or contradict) our results under experimental conditions closely similar to those used here. If our results were corroborated, then the path would be open to investigate whether this effect could be scaled to produce useful electrical power. Even if only voltages far below those for residential power were achievable using our effect, such devices might still have practical applications as "batteries" that would require no fuel and could not wear out in the usual sense.

In principle the diameter of the system depicted in Fig. 1 could be miniaturized without decreasing the generated emf. Using Eq. (26), the coefficient in Eq. (25) may be written as $v^3 (B_\infty/4) l (\sigma \mu a)^2 [1 - (a/b)^2]^{-1}$. Decreasing $b$ and $a$ while increasing the values of $\sigma$ and/or $\mu$ (were materials with such characteristics available or created) or $v$ (for example, in an orbiting satellite) would allow many such devices to be physically placed in parallel but connected in series, amplifying the voltages generated. The emf could also be raised by increasing $l$ or by allowing $a \to b$ in the shell design. We have speculatively considered other configurations satisfying Eqs. (13) and (16) that could produce electric power [2]. Whether scaling one of these examples, or the configuration demonstrated here, could lead to a practical device is a question for further investigation.

## ACKNOWLEDGMENTS

We thank M. Northrup and National Magnetics Group Inc. for producing M100 and other ferrite materials to our specifications and responding to technical questions, and Magnetic Shields Ltd. for producing our MuMetal cylindrical shell. We thank R. J. Wijngaarden, C. Rovelli, R. L. Garwin, P. J. Thomas, and R. Dalal for discussions and G. Z. McDermott, R. D. Holt, and M. L. Lancefield for additional assistance. C.F.C. acknowledges research funding from Princeton University. The views expressed in this paper do not necessarily represent the views of the U.S. government. K.P.H. acknowledges support through the Jet Propulsion Laboratory, California Institute of Technology, under a contract with the National Aeronautics and Space Administration. T.H.C.'s professional affiliation is provided for identification purposes only.

## DATA AVAILABILITY

The data for Figs. 3–7 are openly available [49].

## APPENDIX A: DETAILED DERIVATION OF THE PREDICTED EMF

Chyba and Hand [2] solved Eq. (15) for the magnetic field $\mathbf{B_s}$ in the interior ($a < \rho < b$) of the cylindrical shell for the system shown in Fig. 1. The $x$ component of that solution is Eq. (18). Here we show in detail how this equation leads to the predicted emf.

The emf along some path $\bar{C}$ between the two points $d$ and $e$ within the cylindrical shell is $\langle \varepsilon \rangle_{\rho,\phi}$ given by averaging over $\phi$ and $\rho$ in Eq. (21). For an infinite cylindrical shell, the integrand in this equation for the shell is just equal to [2]

$$\eta \nabla \times B_s = -\nabla V - \eta \nabla^2 \mathbf{A_s} = -\nabla V - vB_{sx}\hat{\mathbf{z}}. \tag{A1}$$

Therefore, Eq. (21) becomes

$$\varepsilon = -\int_{\bar{C}} (\nabla V + vB_{sx})dz. \tag{A2}$$

The condition $R_m < 1$ corresponds to $k\rho < 1/2$ in $B_{sx}(a < \rho < b)$, which may then be expanded by taking series expansions in $k\rho$ (corresponding to powers of $R_m$) for $e^{k\rho}$, $K_0(k\rho)$, and $K_1(k\rho)$ [50]. We have

$$K_0(k\rho) = -\gamma - \ln(k\rho/2) + (k\rho/2)^2[(1-\gamma) - \ln(k\rho/2)] + O(k\rho)^4, \tag{A3}$$

$$K_1(k\rho) = (k\rho)^{-1} + (k\rho/2)[\gamma - 1/2 + \ln(k\rho/2)] + (1/2)(k\rho/2)^3[\gamma - 5/4 + \ln(k\rho/2)] + O(k\rho)^5, \tag{A4}$$

and

$$e^{ky} = 1 + k\rho\sin\phi + (1/2)(k\rho)^2\sin^2\phi + (1/6)(k\rho)^3\sin^3\phi + O(k\rho)^4. \tag{A5}$$

We may now write $B_{sx} = B_{0x} + B_{1x} + B_{2x} + B_{3x} + O(R_m)^4$ as a kind of perturbation series about $B_{0x}$, with successive terms scaled by successive powers of $R_m$. For the zeroth-order term, we find

$$B_{0x}(a < \rho < b) = \beta_1 - \beta_2(a/\rho)^2\cos 2\phi, \tag{A6}$$

which is just the usual field inside a magnetically permeable shell of zero velocity, so that using [2]

$$\nabla V = -v\beta_1\hat{\mathbf{z}}, \tag{A7}$$

we have

$$\nabla V + vB_{0x}(a < \rho < b) = -\beta_2(a/\rho)^2\cos 2\phi. \tag{A8}$$

Because the integrand in the emf integral is independent of $z$, we may average the integrand over $\phi$ (and $\rho$) before performing the integral over $z$. For the lowest-order term we find

$$\langle \nabla V + vB_{0x}\rangle_\phi = \frac{1}{2\pi}\int_0^{2\pi}(\nabla V + vB_{0x})d\phi = 0. \tag{A9}$$

The first-order term, using $k = R_m/2b$, is

$$B_{1x}(a < \rho < b) = -R_m\beta_2 a^2(b\rho)^{-1}\sin\phi\cos^2\phi, \tag{A10}$$

so that

$$\langle B_{1x}\rangle_\phi = -R_m\beta_2 a^2(2\pi b\rho)^{-1}\int_0^{2\pi}\sin\phi\cos^2\phi\, d\phi = 0. \tag{A11}$$

But the second-order term is

$$B_{2x}(a < \rho < b) = -(\beta_2/8)R_m^2(a/b)^2[\ln(k\rho/2) + 4\sin^2\phi - 2\sin^4\phi + \gamma - 1/2], \tag{A12}$$

which does not average to zero over $\phi$ and $\rho$:

$$\langle B_{2x}\rangle_{\rho,\phi} = \frac{1}{\pi(b^2-a^2)}\int_0^{2\pi}\int_a^b B_{2x}\rho d\rho d\phi = -\frac{\beta_2}{8}R_m^2\left(\frac{a}{b}\right)^2\left[\ln\left(\frac{R_m}{4}\right) - \frac{a^2}{b^2-a^2}\ln\left(\frac{a}{b}\right) + \gamma + \frac{1}{4}\right]. \tag{A13}$$

For the system used in this paper, we find $\langle B_{2x}\rangle_{\rho,\phi} = 137$ nT. The third-order term, using $\cos 2\phi = 1 - 2\sin^2\phi$, is

$$B_{3x}(a < \rho < b) = -(\beta_2/8)R_m^3(a/b)^2(\rho/b) \times \{[\gamma + (1/2)\ln(k\rho/2)]\sin\phi + (7/6)\sin^3\phi - (1/3)\sin^5\phi\}, \tag{A14}$$

so that

$$\langle B_{3x}\rangle_\phi = 0. \tag{A15}$$

Equation (25) then follows. We have verified that the fourth-order term $B_{4x}$ does not integrate to zero but makes a negligible contribution to Eq. (21) compared to that of $B_{2x}$.

## APPENDIX B: MATERIALS

The two M100 MnZn ferrite cylinders used in the experiments described in the text were produced to our order by National Magnetics Group Inc. as solid rods through their usual process of sintering. National Magnetics drilled out the long axis of one of these cylinders to produce the cylindrical shell with our requested $a/b$ ratio of 0.6. We determined final actual values for $a$ and $b$ using a digital micrometer, finding $a/b = 0.61$ with $b = 1.0$ cm. Experiments described in the main text measured voltages or currents generated by this cylindrical shell as a function of its orientation. Performing simple regressions over the data for emf and current at the orientations of 0° and 180° yields $|\langle\varepsilon\rangle_{\rho,\phi}| = 17.3 \pm 1.5\,\mu$V and $|\langle i\rangle_{\rho,\phi}| = 25.4 \pm 1.5$ nA, giving a resistance for the shell of $R = 683 \pm 74\,\Omega$. We then calculate $\sigma$ for the shell from

$$\sigma = \frac{1}{R}\int_0^{0.279\mathrm{m}}\frac{dl}{\pi(b^2-a^2)}, \tag{B1}$$

with $b = 0.01$ m and $a = 0.61b$. This gives $\sigma = 2.07 \pm 0.22$ S m$^{-1}$ for the shell. This may be compared with the approximate value for M100 material given in the M100 material data sheet as ≈5 S m$^{-1}$ [35]. (The approximation sign here is because the exact value is not measured in a given production run and has substantial uncertainty; M. Northrup, personal communication.)

The relative permeability $\mu_r$ for M100 material is temperature-dependent, as given by a permeability versus temperature curve presented in the M100 data sheet [35]. The data sheet states that there is a ±30% uncertainty in permeability values for a given M100 sample [35]; we adopt that uncertainty here. Our experiments underlying Figs. 3–5 were conducted over a temperature range of roughly 16 to 22 °C, but those experiments that led to a determination of the emf due to our predicted effect (those at 0° and 180° orientations) took place over a more limited temperature range. As measured by the attached thermocouples, cylindrical shell temperatures in these experiments varied from 18.9 °C to 21.5 °C. The mean temperature determined for all the temperature readings taken for these experiments was 20.0 °C, corresponding to a relative permeability of $\mu_r = 9.5 \times 10^3$, determined by digitizing the relevant plot from the M100 data sheet. With the ± 30% data-sheet uncertainty, we use $\mu_r = 9,500 \pm 2,850$ in calculating $R_m$ and $\langle \varepsilon \rangle$.

Cylinders or cylindrical shells were attached using nonconductive painter's (masking) tape to a rotatable plexiglass turntable on a wood base (that is, no conducting or magnetizable materials were used). The rotation axis of the plexiglass coincided with the origin of an underlying polar coordinate system (affixed to the wooden base) with angles marked in single degrees. The turntable and underlying base were oriented and tilted to be perpendicular to $B_\infty$. At Princeton's location (where our experiments were performed), $B_\infty =$ 45 μT, pointed downward into Earth's surface at an angle (from the horizontal when facing the north geographic pole) of 57.5°, and (in 2022–2024 when the experiments whose results are presented in the text were performed) the direction to the geographic pole was 12.6° east of magnetic north [33]. The turntable was oriented using a SmartTool digital level (with precision 0.1°, but we estimate an uncertainty ±2° due to possible systematic errors) and (far from the ferrite) an analog compass redundantly with the NOAA magnetic field calculator mobile digital compass with precision 0.1° [33], but we estimate an uncertainty ±5° due to possible systematic errors.

The MuMetal cylindrical shell used for our $R_m \gg 1$ control experiment was 15 cm in length, had an outer radius $b = 1.65$ cm, and an inner/outer radius ratio $a/b = 0.75$, and values of $\sigma \approx 2 \times 10^6$ S m$^{-1}$ and $\mu_r \approx 10^5$ [51,52]. This gives $R_m \approx 10^6 \gg 1$ for this cylindrical shell.

## APPENDIX C: AMBIENT LABORATORY ENVIRONMENT

Our experiments were performed in a dark windowless basement room with no heating or other climate control elements present. The only sources of illumination typically present were the LCD displays of the three digital voltmeters and one 60 Hz magnetometer (described in subsequent sections). Care was taken to ensure that even these faint displays had no line-of-sight path to the M100 cylinder or shell. Any possible photoelectric contribution to the experimentally measured voltage was therefore excluded. Investigators operated by headlamp on those unusual brief occasions when they had to be physically present in the room.

We measured ambient temperatures within the laboratory in the proximity of the experimental system using two data-logging Extech RH520 meters (with precision 0.1 °C). The displays of these devices would go dark and remain so for the course of their data-logging. The temperature offset between the two meters had been previously determined by placing their two temperature probes directly adjacent to one another and recording temperatures simultaneously for 6 h. The two probes were then used to investigate ambient vertical and horizontal temperature differences in the laboratory in the vicinity of the experimental system. Vertical temperature gradients were always at least 0.6 °C m$^{-1}$, and could sometimes reach 2 °C m$^{-1}$. Ambient horizontal temperature gradients were typically below 1 °C m$^{-1}$. Ambient horizontal temperature gradients appeared to be due to distance from the laboratory door, as well as distance from walls: two laboratory walls were exterior walls, whereas two bordered interior rooms, leading to different temperatures on opposite sides of the laboratory. We attribute the vertical temperature gradient simply to heat stratification. Our preliminary investigations showed that temperatures inside insulating containers quickly acquired a gradient identical to the ambient room gradient, so vertical gradients were close to unavoidable. Ambient horizontal and vertical gradients of these magnitudes meant that there would be Seebeck voltages in the range $\sim \pm 100$ μV in our system unless additional steps were taken to suppress these ambient temperature differences. Rather than attempt to eliminate these differences we controlled for them via the approach summarized by Fig. 2, using thermocouples attached to the ends of the M100 shell (see Appendix D).

Throughout each run, we monitored 60 Hz magnetic flux density background ($B_{60}$) in the proximity of our experimental system using a Magnii Technologies DSP-523 rms 3-axis meter. The meter recorded the maximum value of 60 Hz (plus harmonics) over the course of each run. The zero of the DSP-523 had been previously determined from the reading of the instrument while deep inside three nested Magnetic Shield Corporation MuMetal cylinders with lids, with a small axial hole in the lids at one end of all three cylinders, allowing observation of the display of the meter at the far end.

For nearly all runs, the 60 Hz signal experienced during the run was about $6 \pm 1$ nT, but with occasional spikes as high as $10 \pm 1$ nT. By taking care to minimize area enclosed by our voltmeter leads, we limited the maximum AC voltage that could be inadvertently induced to $V_{\mathrm{AC}} \approx 6\omega B_{60}\Sigma$, where $\Sigma$ is the cross-sectional area of our cylindrical shell. Approximately 2/3 of $V_{\mathrm{AC}}$ arise due to concentration of flux density by the permeable material, with about 1/3 of $V_{\mathrm{AC}}$ coming from area enclosed by DVM leads outside the footprint of the shell. With $\omega = 2\pi \times 60$ Hz, $B_{60} = 10$ nT, and $\Sigma = 2\,\mathrm{cm} \times 30\,\mathrm{cm}$, $V_{\mathrm{AC}} \approx 0.1$ μV, or two orders of magnitude too small to generate our measured voltages, even if the induced AC voltage were somehow converted to DC voltage with 100% effiency. In fact, we have anecdotal evidence that strong ambient 60 Hz signals suppress the emf measured in our experiments. We suspect that this is because 60 Hz acts as a demagnetizing field, and demagnetization of our magnetically soft cylindrical shell should randomize its magnetic domains and lessen or eliminate our predicted effect. We note that while $B_\infty =$ 45 μT, the generation of the emf is due to the term $\langle B_{2x} \rangle_{\rho,\phi} =$ 137 nT. We suspect that the presence of 60 Hz background signals that approach this value could substantially decrease the generated voltage due to demagnetization effects at the level of the causative perturbation term.

It was shown in the 1960s that ferromagnetic conductors could rectify microwaves into DC signals [53]. Early experiments found conversion efficiencies (DC voltage generated in ferromagnetic sample versus incident microwave power) of ∼1 nV/mW [54]. Subsequent research and development has raised these conversion efficiencies to values as high as 1–100 μV/mW [55,56]. Using an RF Explorer 6G W+ spectrum analyzer, we measured ambient microwave energies in our laboratory across the range 15 MHz to 6.1 GHz and found our highest peak to have a power 0.1 nW. Even if our device somehow rectified microwaves with an efficiency 100 μV/mW, this would generate a voltage of only ∼10 nV, a factor $\sim 10^3$ times smaller than our measured voltages. Moreover, the effect would have to act for the cylindrical shell but not for the solid cylinder, nor for the $R_m \gg 1$ cylinder, and coincidentally have to be such that zero DC signal were generated for the shell at orientations of 90° and 270°.

We have verified that the predicted effect found in our experiments disappears when the experiments are conducted in an $R_m \gg 1$ Faraday cage (we used an LBA Technology Faraday cage made of solid 1/8 inch (0.318 cm)-thick aluminum, giving $R_m \approx 50$). This is expected since in the laboratory frame $K'$, $\mathbf{v} = \mathbf{0}$ so there is no nonzero $q\mathbf{v} \times \mathbf{B}$ force as such. Rather, the predicted effect as seen in $K'$ is driven by the corresponding electric field $\mathbf{E}' = \mathbf{v} \times \mathbf{B}$ [2]. The Faraday cage cancels this field in its interior, thereby also eliminating the effect. The same issue arises inside a room whose walls, ceiling, and floors are predominantly conducting, as would be the case with surrounding steel construction, for example. While our laboratory had some steel elements present, its walls were made of cinder block and its floor of concrete (albeit with rebar present).

The Faraday cage also provides a second argument that our results are not somehow produced by 60 Hz background. The skin depth for magnetic flux density in the walls of the Faraday cage is $\delta = \sqrt{(2/\mu_0 \sigma \omega)} = 1.1$ cm at $\omega = 2\pi \times 60$ Hz and taking $\sigma = 3.5 \times 10^7$ S m$^{-1}$ for aluminum. Therefore, for the Faraday cage with wall thickness 0.318 cm, 60 Hz signals are reduced by a factor $\exp(-0.318/1.1) = 0.75$ in penetrating the walls. Yet our effect entirely disappears within the cage; it does not merely undergo a slight attenuation.

## APPENDIX D: INSTRUMENTATION AND EXPERIMENTAL METHODS

Voltages and temperatures were recorded using three Gossen Metrawatt Metrahit 30M digital voltmeters [labeled by us in Fig. 1 and elsewhere as DVM3 for the voltage (or currrent) measuring device and DVM1 and DVM2 for the temperature measuring devices]. These were battery operated to eliminate any possible 60 Hz effects due to the meters themselves.

The Metrahit 30M provides voltage measurement precision of 0.1 μV (giving a scale error below for inclusion in calculating measurement errors of ±0.05 μV). The DVM3 voltage leads were twisted and attached to the north and south ends of the cylindrical shell using simple crocodile (alligator) clips. Shell surfaces were first prepared by sanding with a Dremel tool to eliminate any oxidation layer that might be present. A system orientation of 0° was defined to be when the cylinder or shell was orientated as in Fig. 1 and the north end of the cylinder or shell was connected to the positive terminal of DVM3. The positive terminal was connected to a position $\phi \approx 145^\circ$ around the shell (see Fig. 1 for cylindrical and Cartesian coordinate definitions), and the negative terminal to a position $\phi \approx 35^\circ$. The distance $l$ for this system was $l = 27.9$ cm.

Each experimental run for a particular orientation of a cylinder or cylindrical shell was begun by rotating the shell to its proper position and then allowing the system to sit overnight. The cylinder and cylindrical shell used in our experiments (apart from the Mumetal control) are composed of MnZn ferrite, a soft magnetic material [34]. Following any given orientation change, a shell's magnetic domains reorient in response to $B_\infty$. There also is likely to be some settling time for measured voltages due to eddy (Foucault) currents created by rotation of the ferrite.

For both the cylindrical shell and the solid cylinder, voltmeter leads were fixed to the rotatable plexiglass turntable, and were rotated together with the shell or cylinder. The voltmeters also moved with the rotation. That is, care was taken to ensure that there was no circuit-topology change under rotation. This was done because it is known that in certain AC systems with $\partial \mathbf{B}/\partial t \neq \mathbf{0}$, potential difference and voltage are distinct quantities, and the measured voltage can vary if the topology of the circuit containing the voltmeter leads varies [38,57]. Because our circuit topology does not change under rotation, this effect cannot be a factor in our measurements.

Measurements were typically begun in the morning, following a period of at least 8 h subsequent to the shell's rotation into position. Data were recorded on the three 30M meters every 10 s for 5–10 h periods. The leads were then disconnected from the Metrahit 30M meter (but always left undisturbed on the ferrite) and data downloaded from each meter to a Panasonic CF-29 Toughbook using Gossen's BD232 interface and software. Care was taken to disturb the ferrite shell or attached leads as little as possible. The 30M meters were powered with Energizer Ultimate lithium batteries, both because lithium batteries have a far more horizontal power versus time curve than other batteries, and because of the much longer life they allow between battery changes, extending to over 30 h of data logging with the 30M meter (though in all cases reported here, batteries were replaced after at most 18 h of use). Replacement batteries were stored near the meters, to keep their temperatures close to those of the meters and minimize any thermal excursions upon battery replacement.

The 30M multimeter DVM3 had a small zero-voltage offset. To determine this offset we created a simple voltage divider, producing nominal voltages of about 16 μV. We then recorded voltage data for a period of 6 h, then reversed the polarity of the leads into the DVM3 terminals, and recorded data for another 6 h. We repeated this procedure for voltages of 32 μV and 48 μV, intending to cover the likely range of voltages to be measured in our actual experiment. Averaging these results together, we found a systematic offset upon reversing polarity of $1.20 \pm 0.35$ μV. Voltage data reported in this paper have taken this offset into account by subtracting 0.60 μV from all recorded voltages.

Similarly, we determined the zero-current offset for the multimeter DVM3 when operating as an ammeter, on the basis of a 9 h data logging run across a 330 ohm resistor. We found an offset of $0.777 \pm 0.053$ nA. All current data reported in this paper have taken this offset into account, by subracting 0.777 nA from all recorded currents.

Temperatures at opposite ends of the cylinder or cylindrical shell were measured using Pt1000 thermocouple probes attached to two Gossen Metrawatt Metrahit 30M battery-operated meters (designated DVM1 and DVM2 in Fig. 1; DVM1 connects to the thermocouple attached at the same end of the cylinder or shell connected to the positive voltage terminal of DVM3). The Metrahit 30M provides temperature measurement precision of 0.01 °C (giving a scale error below for inclusion in calculating measurement errors of ±0.005 °C).

The probes were positioned at $\phi = 180^\circ$ at each end of the cylinder or shell (see Fig. 1), with the distal end of the thermocouple container flush with the end of the cylinder or shell. Since this container is metallic, the thermocouple was separated from the M100 material itself by a (thin) piece of masking tape so that there could be no electrical contact creating an alternate currrent path. Each thermocouple was secured using masking tape over the top of the shell. Care was taken to treat the two thermocouples identically in positioning and attachment. Different pairs of Pt1000 probes were used for the M100 cylinder and cylindrical shell so that once attached, the probes need never be detached from the cylinder or shell.

Because the difference $\Delta T = T(\text{DVM1}) - T(\text{DVM2})$ in temperatures measured by DVM1 and DVM2 is important for our purposes, the relative temperature calibration of DVM1 and DVM2 is crucial. For a given pair of thermocouple probes, prior to their attachment to the cylinder or cylindrical shell, this calibration was determined (after first cleaning the exterior container of each probe) by taping the thermocouple container ends of the Pt1000 probes together and recording temperature data on DVM1 and DVM2 for approximately 6 h. The data were then downloaded and $\Delta T = T(\text{DVM1}) - T(\text{DVM2})$ calculated. For the pair of probes used for the cylindrical shell, this was $\Delta T = 0.012 \pm 0.0065$ °C. For the pair of probes used for the solid cylinder, this was $\Delta T = 0.044 \pm 0.0054$ °C. All $\Delta T$ data reported in this paper have taken these offsets into account. Absolute temperature values in our experiments ranged between 16 °C and 21 °C.

To create plots for experimental data corresponding to the theoretical plot of Fig. 2, our hope with any particular experimental run would be for $\Delta T$ to vary due to variations in the ambient temperature gradients in the laboratory across the range $\Delta T = -0.15$ °C to $\Delta T = +0.15$ °C. Some variation in $\Delta T$ would typically occur due to diurnal or secular temperature variations, but often the range in $\Delta T$ that would be sampled as a result would be only a fraction of the entire range from $\Delta T = -0.15$ °C to $\Delta T = +0.15$ °C. The data for a given orientation in most figures therefore typically represent data from several different runs that, due to changes in ambient conditions of some tenths of a degree, sampled slightly different ranges of $\Delta T$. We also sometimes used ice contained in sealed plastic bags, typically placed between 0.5 m and 1.0 m from either the DVM1 or DVM2 end of the cylinder or cylindrical shell, as necessary, to gradually manipulate the ambient temperature gradient in the laboratory to create conditions approaching the desired range in measured $\Delta T$ values.

The same theory that predicts emf generation for conducting ferrite shells when $R_m \leqslant 1$ also predicts that emf = 0 for a conducting magnetically permeable shell with $R_m \gg 1$ [2]. We tested and verified this prediction with a MuMetal cylindrical shell (see Appendix B). Using this shell we duplicated the south-north orientation and DVM3 electrode positions used in our ferrite shell experiments, recording data every 10 s over a 9.5 h run.

## APPENDIX E: DATA REDUCTION

In every case, including for calibration runs, we dropped the first hour of recorded data from our data to be analyzed, to exclude any effects on measured values due to warm-up time for the 30M meters after they were turned on and data-taking begun. In addition, each 30M meter would experience occasional data drop-outs, always in the first 1.5 h of data-taking during a run; when these occurred in the data of any one meter at a particular time, we would drop the data for all three meters for that time from our analysis.

The plots shown in Figs. 3–7 were produced by grouping all data measured for a particular orientation of the system by $\Delta T$ value using a simple matlab code. The mean of the corresponding N emf values then gives the data point shown for that $\Delta T$ value. The half-height of the error bar for each mean emf value is computed as the pythagorean sum of the sample standard deviation for that point (i.e., for that $\Delta T$ value) and the scale error of ±0.05 μV. The half-width of the error bar for each $\Delta T$ value is simply the relevant scale error of ±0.005 °C.

Figure 6 compares results for emf as a function of temperature gradient for the solid $a = 0$ cylinder, after combining data for all four orientations shown in Fig. 4, with those for the $a/b = 0.61$ cylindrical shell for orientations 90° and 270° combined. The points are colinear (albeit with some scatter), as predicted. The slope of a simple linear regression fit to the $a = 0$ cylinder gives a Seebeck coefficient $S = -417$ μV $\text{K}^{-1}$.

## APPENDIX F: SECOND EXPERIMENTAL SITE

As an additional check to test whether an unrecognized local effect could somehow be spoofing the array of predicted results verified in our primary laboratory, we reproduced our experiments in a largely uncontrolled environment in a residential building 5.5 km east of our laboratory's location. We conducted these "Site B" experiments in a dark walk-in closet on the top floor of a private home. The closet was often subject to rapid temperature fluctuations due to uncontrolled air conditioning during the months of July and August 2024. The 60 Hz background in the closet at Site B was higher than in our primary laboratory, with typical values of $B_{60} =$ 20 nT and fluctuations reaching 35 nT. We measured ambient microwave energies in the closet across the range 15 MHz to 6.1 GHz and found our highest peaks to have a power ~1 nW, about an order of magnitude stronger than had been the case in our primary laboratory. But similarly to that case, even if our device somehow rectified microwaves with an efficiency 100$\mu$V/mW, this would generate a voltage of only ~100 nV, a factor ~$10^2$ times smaller than our measured voltages. Moreover, this effect would again coincidentally have to be such that no DC signal was generated for the shell at orientations of 90° and 270°, and this coincidence would have to hold both in our laboratory and at Site B.

Experimental data for Site B are noisy with correspondingly larger error bars in comparison with the results obtained in our primary laboratory. Nevertheless, the data once again show voltage magnitude and behavior consistent with our predictions, as seen in Fig. 7. These results imply that the data obtained in our primary laboratory are not due to an unidentified effect that is somehow spoofing the behavior predicted for our effect.

Ambient room temperatures at Site B typically lay between 23 °C and 25 °C. The mean temperature over all of our data taken at orientations of 0° and 180° was 24.0 °C. This leads, according to the permeability versus temperature curve included in the M100 datasheet [35], to a higher relative permeability for the M100 shell than had been the case in our laboratory (where, as noted above, the mean temperature over all data taken at orientations of 0° and 180° had been 20.0 °C). Moreover, it has been shown that the conductivity of MnZn ferrites also increases with increasing temperature, as long as one remains below the Curie point [36]. Since $\langle\varepsilon\rangle$ varies roughly like $\mu_r^2\sigma^2$, we expect our experiment to generate somewhat higher emf values at Site B than in our primary laboratory. This is observed, but the size of the one-sigma error bars for emf measurements in this uncontrolled environment makes us reluctant to attempt too much quantification of these results.

Because at Site B our multimeters were operating in a warmer environment than had been the case in our primary laboratory, we determined anew the voltage offset for DVM3 and the offset in $\Delta T$ between DVM1 and DVM2. We reevaluated the emf offset for DVM3; we found emf = 0.27 ±0.10 µV over a 15 h run. The voltage data shown in Fig. 7 have taken this slightly changed offset into account by subtracting 0.27 µV from all recorded DVM3 voltages. We also reevaluated the $\Delta T$ offset for DVM1 and DVM2; we found $\Delta T = 0.0026 \pm 0.0072$ °C over a 14.5 h run. The data shown in Fig. 7 have taken this offset into account by subtracting 0.0026 °C from all $\Delta T$ values.

*Correction:* Equations (24), (25), and (A13) contained minor errors and have been fixed.